\documentclass[]{pasj02} 
\usepackage[switch,mathlines]{lineno} 
\usepackage{aas_macros}
\usepackage{natbib} 
\usepackage{bm}
\usepackage{appendix}

\jyear{2026}
\Received{2026/08/05}
\Accepted{2028/10/28}

\begin{document} 

\title{photo-$3\times2$-pt: Cosmology from cosmic shear and
  galaxy clustering with a single photometric galaxy catalog}

\author{
  Takashi \textsc{Hanama}
}
\altaffiltext{}{National Astronomical Observatory of Japan, Mitaka, Tokyo 181-8588, Japan}

\KeyWords{cosmology: observations — dark matter — cosmological
  parameters — large-scale structure of universe}

\maketitle

\begin{abstract}
  We perform a joint cosmological analysis using angular power spectra
  of galaxy clustering, galaxy-shear cross correlation, and cosmic shear 
  (hereafter referred to as photo-$3\times 2$-pt) measured from
  the Hyper Suprime-Cam year-3 (HSC-Y3) weak lensing shape catalog. 
  We employ the pseudo-$C_\ell$ method to measure these power spectra
  and use the template deprojection method to mitigate systematics in
  the galaxy density maps.
  We perform a standard Bayesian likelihood analysis for cosmological
  inference based on the measured photo-$3\times 2$-pt, including
  contributions from intrinsic alignments of galaxies, galaxy clustering
  bias, lensing magnification effect, baryonic feedback effect, and
  source redshift distribution errors.
  For a flat cold dark matter model, we find a 68\% credible interval of
  $0.76 \le S_8 \le 0.81$ ($S_8=\sigma_8 \sqrt{\Omega_m/0.3}$).
  This result is consistent with the HSC-Y3 cosmic shear analyses
  \citep{Dalal_2023,Li_2023} but is $\sim 25$\% tighter.  
  We also find that the measured B-mode power spectra are consistent
  with zero.
  We also perform a performance test using mock catalogs that mimic the
  HSC-Y3 data to validate our photo-$3\times 2$-pt analysis.
  Through Bayesian parameter inference of mock data, we verify that
  unbiased estimates can be obtained for the key cosmological and
  nuisance parameters, including $S_8$, intrinsic alignment model
  parameters, galaxy clustering bias, and shift
  parameters of the source redshift distributions.
\end{abstract}

%
%
\section{Introduction}
\label{sec:intro}

Cosmic shear two-point correlation functions or the power spectra
depend both on the time evolution of the cosmic structure and on
the cosmic expansion history at relatively recent epochs
($z < 1$), and thus serve as a unique and most powerful late-time
cosmological probe.
With the aim of placing useful constraints on cosmological
parameters independently of early-time probes
\citep[e.g.,][]{2020A&A...641A...6P}, three stage-III
cosmological surveys, namely Dark Energy Survey (DES,
\citet{2016MNRAS.460.1270D}), Kilo-Degree survey (KiDS,
\citet{kids_2013}), and Hyper Suprime-Cam Subaru Strategic Program (HSC
survey, \citet{2018PASJ...70S...4A}), were conducted and have completed
their observations.
DES and KiDS have published cosmic shear results from their final data
\citep{https://doi.org/10.48550/arxiv.2602.10065,Wright_2025}, yielding
constraints with below 2\% precision on
$S_8=\sigma_8(\Omega_m/0.3)^{0.5}$, where $\sigma_8$ is
the amplitude of matter fluctuations in scales of 8Mpc/$h$
and $\Omega_m$ is the mean matter density parameter.
HSC survey team is currently carrying out cosmic shear analyses using the
final data. 
Now with the start of science operations of the Euclid
\citep{2011arXiv1110.3193L} and the Rubin LSST
\citep{2009arXiv0912.0201L}, weak lensing surveys are entering
the stage-IV. 

Observing a larger volume is a primary approach to reducing errors in
cosmological parameter constraints.
Indeed, this is currently taking
place with the transition from the Stage-III to Stage-IV weak lensing
surveys.
At the same time, it is equally crucial to develop methods for
extracting the maximum amount of cosmological information from a given
survey dataset.
This is exactly the purpose of this study.

We use the HSC-survey year 3 (HSC-Y3) weak lensing shape catalog
\citep{Li_2022}, which was used in the cosmic shear 
analyses by \citet{Dalal_2023,Li_2023}.
We utilize two-dimensional galaxy number density fields of
tomographic weak lensing galaxy samples to extract information contained
therein, in addition to the weak lensing shear fields.
Specifically, we use the galaxy clustering angular power spectra and
galaxy-shear cross power spectra, in addition to the shear-shear (cosmic
shear) power spectra.
As discussed in section~\ref{ssec:constraingpower}, these new additions
provide useful information particularly on intrinsic alignment signals
in galaxy shapes and redshift distributions of galaxy samples
\citep{2011A&A...527A..26J,Johnston_2025}.
These are major nuisance parameters that result in increased errors
in cosmological parameter constraints.
In fact, uncertainties in galaxy redshift distributions are a serious
issue identified in HSC-Y3 cosmic
shear studies \citep{Dalal_2023,Li_2023}.

Our approach is similar to the standard ``$3\times 2$-pt'' analysis
\citep[e.g., ][]{Miyatake_2023,Sugiyama_2023}, which combines
spectroscopic galaxy samples with a weak lensing shape catalog to derive
three types of two-point correlation functions (or power spectra):
galaxy clustering, galaxy-shear cross-correlation (i.e., galaxy-galaxy
lensing), and cosmic shear.
In fact, we employ the same types of 2-pts, but all the
power spectra are
derived from the single photometric galaxy catalog; we refer to this as
``photo-$3\times 2$-pt''.
The primary advantage of this approach is that the additional
information extracted from the galaxy clustering and galaxy-shear
cross-spectra originates from the same galaxy sample used in the cosmic
shear analysis, but arises from the different physical source, the
galaxy density fields.
These power spectra may be affected by nuisance parameters
common to the cosmic shear in different ways.
Therefore the new information can be complementary to
that obtained from the shear fields.
Consequently, the newly added independent information can improve
constraints on common nuisance parameters, potentially yielding tighter
constraints on cosmological parameters than an analysis based solely on
cosmic shear
\citep[see][for previous studies examining similar methods]{Schaan_2020,Johnston_2025}.

The structure of this paper is as follows.
In section \ref{sec:dataset}, we briefly summarize the HSC-Y3 weak
lensing shape catalog and the photometric redshift data used in this
study. 
In section \ref{sec:measurements}, we describe the method to measure the
three types of angular power spectra and present our measurements.
The results of the B-mode null test are also presented.
In section \ref{sec:models}, we present an overview of the
theoretical modeling of 
the three types of power spectra and the covariance matrix.
In section \ref{sec:inference}, we describe a
method for parameter inference along with methods to take into account
various systematics in our cosmological analysis.
There, we define the scale-cut of power spectra
that form the data vector adopted in our analyses, and 
we detail model parameters and their priors.
In section \ref{sec:results}, we present the results of
our cosmological constraints and tests for systematics.
We compare our cosmological constraints with those
from the HSC-Y3 cosmic shear analyses.
Finally, we summarize and discuss our results in
section \ref{sec:summary}.
In appendix \ref{apdx:mock},
we describe a performance test of the photo-$3\times 2$-pt analysis with mock
catalogs.
We first describe numerical simulations to create mock catalogs
for photo-$3\times 2$-pt measurements, then we present results from cosmological
analyses of the mock data.
In appendix \ref{apdx:systematics}, we describe details of area cuts of
galaxy density maps to mitigate effects of systematics on galaxy
clustering measurements.

Throughout this paper we quote 68\% credible intervals
for parameter uncertainties unless otherwise stated.

%
%
\section{HSC-Y3 weak lensing shear catalog}
\label{sec:dataset}

We use the same HSC-Y3 weak lensing shear catalog as that used in
the cosmic shear analyses of \citet{Dalal_2023,Li_2023}, and thus here
we focus on aspects that are directly relevant to this study.
We refer the readers to those two papers and references therein for
details.

The public HSC three-year shape catalog \citep{Li_2022} covers
433 deg$^2$ with an effective galaxy number density of 19.9
arcmin$^{-2}$.
For the cosmic shear analyses, additional cuts by galaxy properties,
conservative masking, and qualities of the photometric redshift
estimation, were applied \citep[see][for details]{Dalal_2023,Li_2023}.
After those cuts, the final shear catalog contains 25 million galaxies
covering 416 deg$^2$.
Survived galaxies were divided into four tomographic redshift bins by
using the {\tt best estimate} by the {\tt dNNz} photo-z algorithm
\citep{2018PASJ...70S...9T}.
The effective number densities of galaxies in each tomographic redshift
bins are \citep{Li_2023} 3.77, 5.07, 4.00, and 2.12arcmin$^{-2}$ for
redshift intervals of (0.3, 0.6], (0.6, 0.9], (0.9, 1.2],
and (1.2, 1.5], respectively.

Galaxy shapes were estimated using the re-Gaussianization ({\tt reGauss})
PSF correction method
\citep{2003MNRAS.343..459H,2005MNRAS.361.1287M,2013MNRAS.432.1544M}
implemented in {\tt GalSim} \citep{Rowe_2015}. 
Those {\tt reGauss} shape measurements were calibrated using deep image
simulations, which reduces the galaxy property-dependent multiplicative
shear estimation bias to less than $\sim 10^{-2}$ \citep{Li_2022}.
\citet{Li_2022} also presented a series of null tests, and
demonstrated that additive biases due to PSF modeling and shear
estimation errors are overall within requirements defined for cosmological weak
lensing sciences using this shape catalog.
They, however, found that the auto-correlation of the PSF model shape
residual marginally exceeds the requirement on $> 1^\circ$ scales.
In HSC-Y3 cosmic shear studies \citep{Dalal_2023,Li_2023}, the PSF
induced additive biases were corrected by the method developed and
thoroughly tested by
\citet{Zhang_2021,Zhang_2022,Zhang_2023}, in which the
second and fourth 
PSF moments, rather than just the second moments, were used to model
the PSF related contributions to the cosmic shear correlations.
In their studies, it is found that the impact of the
PSF additive biases on constraints on $S_8$ parameter is small.
Furthermore, \citet{Jefferson2025Reanalysis} conducted a comprehensive
reanalysis of the Stage-III shape
catalog including HSC-Y3 by performing a series of diagnostic
tests, some of which had not been performed by the HSC collaboration.
They also assessed the impacts of PSF related systematics on the
$\Omega_m$-$S_8$
parameter space by comparing the posteriors of simulated data vectors with
and without PSF contamination, finding the effects to be negligible.

The redshift distribution of four tomographic galaxy samples are
inferred by \citet{Rau_2023}, in which the photometric redshift
estimation and spatial cross-correlation between the HSC-Y3 shape
catalog and the CAMIRA-LRG catalog \citep{Oguri_2014,Oguri_2017} were
jointly used.
We refer a reader to \citet{Rau_2023} for details.
The derived redshift distributions are included in the public HSC-Y3
data products, 
and we use them in our computations of theoretical models.

%
%
\section{Measurements of power spectra}
\label{sec:measurements}
We measure three types of power spectra:
the shear-shear, galaxy clustering, and galaxy-shear power
spectra. 
An overview of the measurements is as follows:
We first generate maps of shear field and galaxy number density field.
Those maps are in {\tt HEALPix} pixelization format \citep{Gorski_2005},
with {\tt HEALPix} parameter of $N_{\rm side}=4096$ or an effective area
of $0.74$arcmin$^{2}$.
We compute the angular power spectra from those maps adopting the
pseudo-$C_\ell$ formalism implemented in {\tt NaMaster}
\citep{Alonso_2019,Nicola_2020,Nicola_2021}, which 
enable to correct for biases and mode-coupling due to a survey geometry.
The power spectra are binned into consecutive band powers
with a constant logarithmic spacing of $\Delta \log_{10}\ell = 0.125$.

In the following subsections, we describe some details of measurements
of three types of power spectra in turn:

%
%
\subsection{Shear-shear power spectra}
\label{ssec:shear-shear}

We estimate shear-shear power spectra using almost the same
procedure as \citet{Dalal_2023}.
The only difference is that, whereas \citet{Dalal_2023} measured power
spectra individually for six disconnected fields using the flat-sky
mode, and then combined them, we employed the full-sky mode. 
In fact, we confirmed that power spectra we measured agree well with
those of \citet{Dalal_2023}.

Here, we only describe key aspects of measurements, 
as we followed the same procedure as \citet{Dalal_2023}. 
We refer the readers to the paper and references therein for details.
In order to compute the shear-shear power spectrum using {\tt NaMaster},
we need to prepare a shear map and its associated weight map
\citep[see][for details]{Nicola_2021} for each tomographic galaxy sample.
The shear map is a pixelized shear field where pixel values
are weighted mean shears of galaxies within each pixel \citep{Nicola_2021}.
The shear of each galaxy can be estimated by applying conversion and
correction factors to measured galaxy ellipticity, all of which are
included in the HSC-Y3 shape catalog \citep{Dalal_2023}.
The weight of each galaxy is also included in the catalog,
and is the inverse variance of the shape noise \citep{Li_2022}.
Pixel values of the weight map are the inverse variance which can be
estimated by summing-up weights of galaxies within each pixel
\citep{Nicola_2021}. 
Having prepared the shear and weight maps for four tomographic
galaxy samples, we compute the auto- and cross-power spectra
for their ten combinations using the pseudo-$C_\ell$ method implemented in
the {\tt NaMaster} software \citep{Alonso_2019}.

%
%
\subsection{Galaxy clustering power spectra}
\label{ssec:g-g}

%
%
\begin{figure}
\begin{center}
\includegraphics[width=82mm]{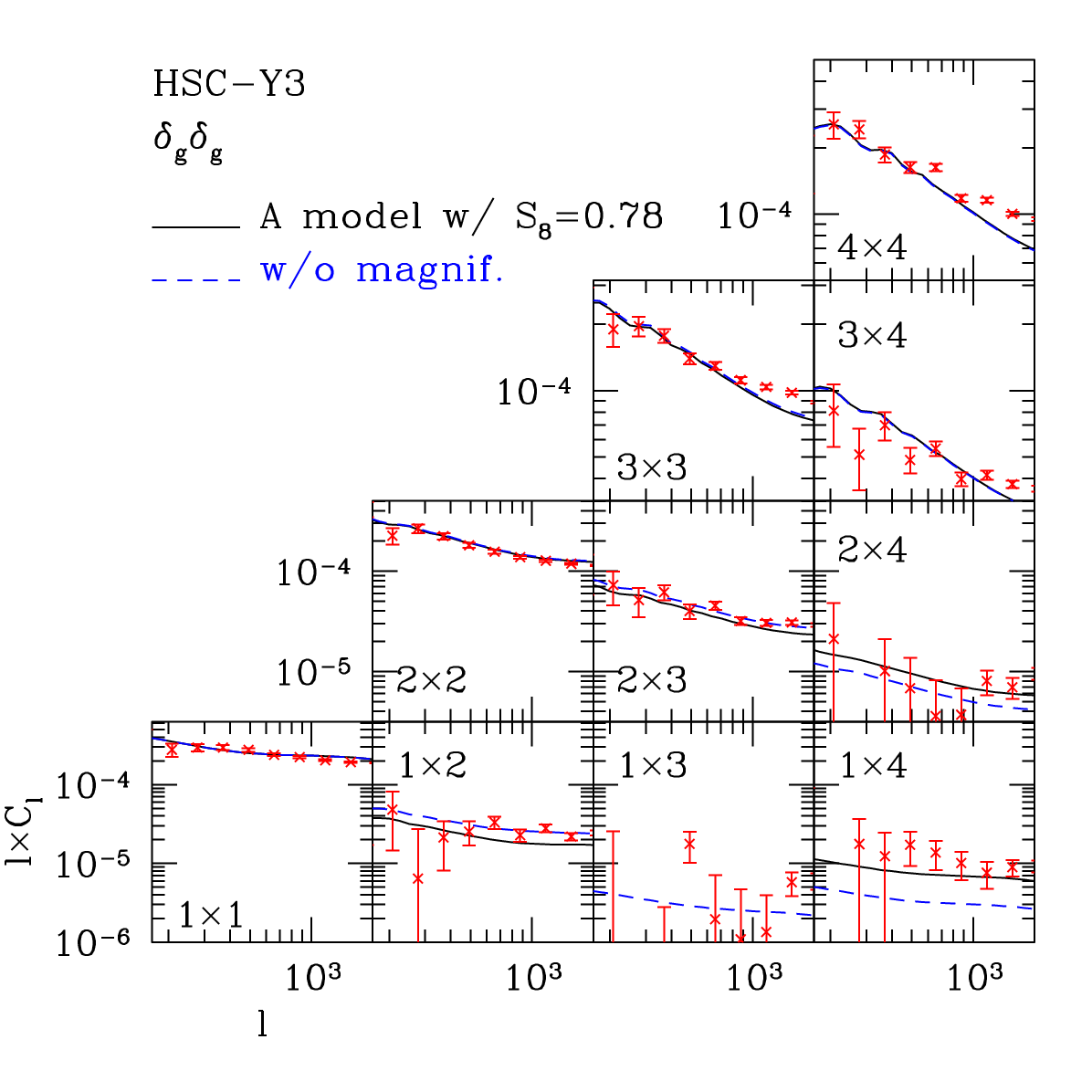}
\end{center}
\caption{Galaxy clustering power spectra for ten combinations of
  tomographic redshift bins (indicated in each panel).
  Error bars represent the square-root of the diagonal elements of the
  covariance matrix.
  Note that only four $\ell$-bins within the range $178 \le \ell \le
  562$ of the auto-power spectra are used for parameter inference.
  The cross-spectra are also presented to show levels of detections.
  The solid lines show the theoretical prediction based on a representative
  $\Lambda$CDM model with $S_8=0.78$, where other model parameters are
  consistent with the result of the fiducial parameter inference.
  The dashed lines are the same theoretical model but without 
  the magnification effect.
  \label{fig:clgg_hsc}}
\end{figure}

We measure galaxy clustering power spectra largely following the
method developed in \citet{Nicola_2020}, in which magnitude limited
photometric galaxy samples from the first public data
release of the HSC survey \citep{2018PASJ...70S...8A} were analyzed.
Its procedure consists of the four steps:
\begin{enumerate}
\item We make maps of survey mask and galaxy number densities.
\item We make maps of systematics that may modify the observed 
distribution of galaxies, and thus may affect their clustering properties.
\item We mitigate effects of systematics in two ways, one is to cut areas
significantly affected by systematics, and the other is to deproject the
maps of these systematics from the galaxy density maps using ``template
deprojection'' method \citep{Elsner_2016} implemented in the
{\tt NaMaster} software \citep{Alonso_2019}.
\item Finally we compute galaxy clustering power spectra using the {\tt
  NaMaster} \citep{Alonso_2019}.
\end{enumerate}
Below we describe each step in turn.

\subsubsection{Survey mask and galaxy number density maps}
The survey mask is a pixel-based record of survey geometry.
We make it using the HSC {$i$}-band random catalog retrieved from the HSC
public database\footnote{\tt https://hsc-release.mtk.nao.ac.jp/} and the
HSC-Y3 shape galaxy catalog itself.
The random catalog contains random points over survey regions with a
number density of approximately 100 arcmin{$^{-2}$}.
We remove random points within the star
masks in the same manner
as one adopted in making HSC-Y3 shape catalog\footnote{Specifically,
random points with any of star 
mask flags with {\tt i\_mask\_barightstar\_ghost15},
{\tt i\_mask\_barightstar\_halo} or 
{\tt i\_mask\_barightstar\_blooming} are removed.} \citep{Li_2022}.
We make the random binary mask (denoted by {\tt rmask}) such that {\tt
  rmask} is 0/1 if there is no/any random point within each pixel.
We also make galaxy binary mask (denoted by {\tt gmask}) in the same
manner, in which all four tomographic galaxy samples are used (the mean
galaxy density of {$\sim 15$} arcmin{$^{-2}$}). 
Since the mean number of galaxies in each pixel is
{$\sim 11$}arcmin{$^{-2}$},
there is a chance that pixels on survey region contain
no galaxy by chance.
Thus if a pixel with {\tt gmask}=0 is surrounded by {\tt gmask}=1 pixels,
we set {\tt gmask} of that pixel to 1.
Note that regions where {\tt rmask}=1 but {\tt gmask}=0 are mostly due
to problems with data in at least one band ($g$-, $r$-, $i$-, $z$-, or
$y$-band), resulting in a cut by the full-depth and full-color
condition \citep{Li_2022}, or insufficient input in photometric redshift
estimation.
Then we conservatively set pixels that contact with any
{\tt gmask}=0 pixel to {\tt gmask}=0. 
Finally we combine the random mask and galaxy mask (denoted by {\tt rgmask})
so that if either mask is 0, we set {\tt rgmask}=0.

Galaxy number density of each pixel is estimated by numbers of
galaxies in each pixel, and fractions of the area covered
by the survey region in each pixel.
The former is estimated by simply counting galaxies in each pixel,
whereas the latter is estimated by using the random catalog as follows:
We count numbers of random points in each pixel.
Then we normalize the counts using the most frequent value, 70, while
setting an upper limit of 1. 
The resulting normalized counts, referred to as {\tt pixel-coverage
  rate}, is an estimate of the effective pixel area, which is used to
estimate the effective galaxy
number densities for each pixel.
The {\tt pixel-coverage rate} is also used as a systematic map in galaxy
clustering analyses.

\subsubsection{Systematics maps}
Systematics maps are {\tt HEALPix} format maps of astrophysical and
observational effects that can affect the observed distribution of galaxies.
We consider the following ten systematics, which are chosen based
  on physical considerations and previous studies
  \citep[e.g,][]{Leistedt_2016,Nicola_2020,Rezaie_2020,Johnston_2021,Rodr_guez_Monroy_2022,https://doi.org/10.48550/arxiv.2509.07943,S_nchez_2023,Yan_2025}
  as the most plausible sources of systematic variations.
\begin{enumerate}
\renewcommand{\labelenumi}{\alph{enumi}.} 
\item {Survey depth:} We generated a map of the 5$\sigma$ point
  source survey depth using the star catalog associated with the HSC-Y3
  shape catalog\footnote{\tt
  https://hsc-release.mtk.nao.ac.jp/doc/index.php/\\
  s19a-shape-catalog-pdr3/}.
  Since the mean number density of stars in the star catalog is 
  $\sim 1.3$ arcmin$^{-2}$, and thus is not sufficient to fill all pixels
  with at least one star, we generate a map with lower resolution
  ($N_{\rm side}=2048$) and then upgrade it to $N_{\rm side}=4096$.
  Even for the lower resolution case, there are empty pixels, for which
  values are interpolated from surrounding pixels.  
\item {Dust extinction:} We generated a map of the Galactic
  extinction in $i$-band using data ({\tt a\_i}) provided from HSC
  database,
  which is derived from the extinction map of \citet{Schlegel_1998}.
\item{Star contamination:} We generated a map of the number density of
  stars using the star catalog associated with the HSC-Y3
  shape catalog. The correction for the {\tt pixel-coverage rate}
  was applied to it.  As in the case of the survey depth map, we first
  generate a map with lower resolution ($N_{\rm side}=2048$) and then
  upgrade it to $N_{\rm side}=4096$. We do not apply the interpolation for
  empty pixels, because zero is actual information for this case.  
\item{Observing condition:} We generated maps of observing conditions,
  which likely affect observed distributions of galaxies.
  We use metadata of CCD visits that go into HSC-Y3 coadded images,
  retrieved from HSC database.
  Based on physical considerations and above mentioned previous
    studies, we consider the following six quantities;
  (1) sky count level, (2) inverse variance of sky count, (3) airmass,
  (4) exposure time, (5) number of visits, and (6) seeing condition.
  All those quantities are for $i$-band, as the HSC-Y3 shape catalog was
  based on galaxies primary detected in $i$-band image.
  We produced maps of those quantities in the same {\tt HEALPix}
  format as other maps.
  For the above (1), (3) and (6), we produced coadded maps by computing
  weighted mean of those values over overlapped visits for each pixel.
  The weight used in the coadd process is the inverse variance of sky
  count. For the above (2) and (4) a simple sum over overlapped visits
  for each pixel is used.
  We also utilize the {\tt pixel-coverage rate} map as a systematic map.
\end{enumerate}

\subsubsection{Mitigating effects of systematics}
\label{sssec:mitigating}
We mitigate effects of systematics in two steps following
\citet{Nicola_2020}, one is to cut areas
most likely affected by systematics, and the other is to deproject the
maps of these systematics from the galaxy density maps.
Details of the former is summarized in
Appendix~\ref{apdx:systematics}, in summary we cut $\sim 30$ percent
of the survey area.
We describe the latter below.

For the second step, we adopt the
``template deprojection'' method developed by
\citet{Elsner_2016,Nicola_2020}, which is implemented in NaMaster
\citep{Alonso_2019}. 
The core concept and procedure of this method are as follows (see the
reference above  for details):
The basic assumption is that contaminants
linearly affect the observed galaxy over-density as
$\delta_g^{\rm obs} =\delta_g^{\rm true} +A_{\rm syst} \Delta_{\rm syst}$,
where $\Delta_{\rm syst}$ is
a template map (systematic map in our terminology) of the fluctuations
of a given contaminant around its
mean over map, and $A_{\rm syst}$ is an unknown linear coefficient.
An algorithm implemented in {\tt NaMaster} determines the
best-fit value of $A_{\rm syst}$, and subtracts the corresponding
best-fit contaminant contribution from the maps.
Finally, the angular power spectrum is computed while analytically
accounting for the associated loss of modes.

This method was adopted by \citet{Nicola_2020}, who analyzed
galaxy samples from the first public data release of the HSC survey, and
rigorously validated its performance.
They demonstrated that the level of contamination in the raw galaxy
density maps is low, and the template deprojection method is
sufficiently accurate to account for it. 
The HSC-Y3 shape catalog used in this study shares very similar
characteristics with the first public data, as both are based on the
same HSC survey data, though the HSC-Y3 covers a wider area.
Therefore, the method is expected to perform equally well
on our dataset.

Although we have produced 10 systematics maps, some exhibit strong
  correlations among them, because some are secondary parameters derived
  from other, more primary ones \citep{Leistedt_2016,Rezaie_2020}. 
  To avoid redundancy, we select 6 contaminants maps based on an
  inspection of their cross-power spectra.
  Selected are the dust extinction, star
  number density, sky count level, seeing, total exposure time, and
  pixel-coverage rate.
  The remaining maps are not selected for the following reasons: survey
  depth is strongly correlated with both total exposure time and seeing;
  the inverse variance of the sky counts is strongly correlated with
  total exposure time; the number of visits provides essentially the
  same information as the total exposure time; and 
  airmass is found to have a negligible correlation with
  galaxy number density after the area cut.
\subsubsection{Power spectrum measurements}
Having prepared the galaxy over-density maps for four tomographic
galaxy samples, the mask, and contaminants maps, we compute four auto
power spectra using the pseudo-$C_\ell$ method implemented in 
the {\tt NaMaster} \citep{Alonso_2019}.
The results are presented in Figure~\ref{fig:clgg_hsc}.
Note that we also compute cross-power spectra for their six
combinations in the same manner as the auto-power spectra, but we do not
use them for the parameter inference (see Section~\ref{ssec:datavec}).

%
%
\subsection{Galaxy-shear cross power spectra}
\label{ssec:g-shear}

%
%
\begin{figure}
\begin{center}
\includegraphics[width=82mm]{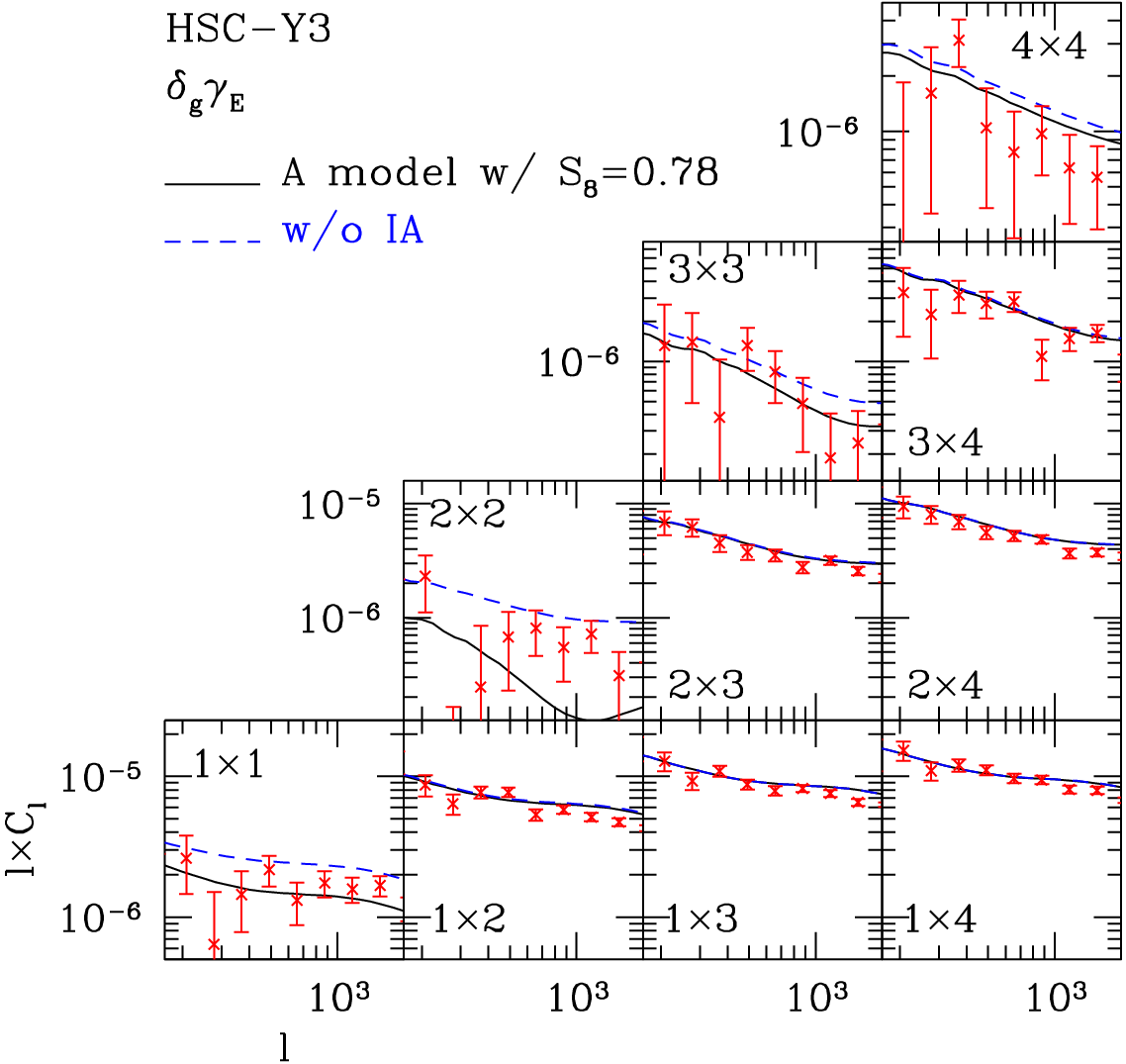}
\end{center}
\caption{Galaxy-shear power spectra for ten combinations of tomographic
  redshift bins (indicated in each panel). Error bars represent the 
  square-root of the diagonal elements of the covariance matrix.
  Note that only four $\ell$-bins within the range $178 \le \ell \le
  562$ are used for parameter inference.
The solid lines show the theoretical prediction based on a
representative $\Lambda$CDM model with $S_8 = 0.78$, where other model
parameters are consistent with the result of the fiducial parameter
inference. The dashed lines are the same theoretical model but without
contributions from intrinsic alignment.
  \label{fig:clge_hsc}}
\end{figure}

Galaxy-shear cross power spectra can be computed using maps prepared for
shear-shear, and galaxy clustering power spectra using the {\tt
  NaMaster} software \citep{Alonso_2019}.
We computed all the 10 combinations of galaxy-shear cross
power spectra with shear maps being at the same redshift bin as or
higher redshift bins than a galaxy map.
The results are presented in Figure~\ref{fig:clge_hsc}.

%
%
\subsection{B-mode null test}
\label{ssec:bmode}

Since in the standard gravity theory with a scalar potential, to
first order, no B-mode signal is expected for the shear-shear and
galaxy-shear power spectra, 
the presence of B-modes may be an indicator of potential 
systematic effects in the data, such as contamination by the
PSF\footnote{B-modes are also generated by second-order lensing
  effects \citep{Krause_2010}, second-order intrinsic alignment
  contributions \citep{Blazek_2019}, the clustering of source galaxies
  \citep{2002A&A...389..729S} and
  the lensing magnification effect \citep{Yu_2015}.}. 
Thus, we quantitatively test the consistency of the B-mode
power spectra signals with zero using the standard $\chi^2$ statistics,
$\chi_B^2 = \sum_{ij} d_{B,i} {\rm Cov}_{B,ij}^{-1} d_{B,j}$
where $d_{B,i}$ is a B-mode data vector,
and ${\rm Cov}_{B,ij}$ is a B-mode covariance matrix consisting only
of the Gaussian contribution from the shape noise
(see section~\ref{ssec:Covariance}).
We take the same $\ell$-range as the parameter inference of E-mode data
(see section~\ref{ssec:datavec}),
namely $317 \le \ell \le 1000$, and $178 \le \ell \le 562$ for the
shear-shear, and galaxy-shear power spectra, respectively. 
Thus the degree-of-freedom (dof) is 40 ($=4$ bins$\times$10 spectra).

For the shear-shear power spectra,  
we find $\chi_B^2=53.5$ ($27.3$) with dof$=40$, corresponding to a
p-value of 0.08 (0.94) for BB-mode (EB-mode).
For the B-mode galaxy-shear power spectra,  
we find $\chi_B^2=35.4$ with dof$=40$, corresponding to a
p-value of 0.68.
Therefore we conclude that no evidence for a significant B-mode signal
is found. 
Note that various other diagnostic tests of PSF-related systematics
  are presented in the HSC-Y3 shape catalog paper \citep{Li_2022}, and in
\citet{Jefferson2025Reanalysis}. We refer the readers to those two
papers and references therein for details (see also Section
\ref{sec:dataset} for a brief summary).

%
%
\section{Theoretical models}
\label{sec:models}

%
%
\subsection{Angular power spectra}
\label{ssec:Angularpowerspecrtra}

Here we summarize expressions for the theoretical models of shear-shear,
galaxy-shear, and galaxy clustering power spectra.
Our theoretical models are based on the following models or assumptions;
the standard $\Lambda$CDM model, the linear tidal intrinsic alignment model
with nonlinear matter power spectrum inserted 
\citep{Hirata_2004,2007NJPh....9..444B,2015SSRv..193....1J}, 
the standard perturbation framework for the matter power spectrum with
upto one-loop terms and with the linear term replaced with the nonlinear
model \citep{Blazek_2015}, and the scale independent linear galaxy
bias models.

The observed shear field consists of two contributions,
\begin{equation}
\label{eq:shearobs}
\gamma_{\rm obs}^i = \gamma^i + (1+\delta_g^i) \gamma_{\rm IA}^i,
\end{equation}
where the superscript $i$ denotes $i$-th tomographic sample, $\gamma$ is
lensing shear, $\delta_g$ is the galaxy density fluctuation field,
and $\gamma_{\rm IA}$ is the contribution from the intrinsic alignment (IA).
The $(1+\delta_g^i)$ factor in front of
  $\gamma_{\rm IA}^i$ arises because IA signals are sampled only at
  galaxy positions, which are generally clustered, rather than being
  uniformly sampled.
Consequently, IA-related correlations are
{\it density-weighted} and thus biased compared to the case of uniform
sampling, depending on 
the relation between the galaxy density field and the tidal
field at observed galaxy positions.
Assuming a scale independent linear bias model, we compute the
density-weighted IA correlation terms by replacing $\delta_g^i$ in
the above equation (\ref{eq:shearobs}) with $b_T^i \delta_m^i$, where
$b_T^i$ is the tidal IA bias parameter and
$\delta_m^i$ is the matter density fluctuation field.
The galaxy density fluctuation field also consists of two
contributions,
\begin{equation}
\label{eq:delta_g}
\delta_{g, \rm obs}^i = b_g^i \delta_m^i + \delta_{\mu}^i,
\end{equation}
where we have assumed a scale independent linear bias model for galaxy
clustering with the bias parameter $b_g$, and $\delta_\mu$ is the 
contribution from the lensing magnification effect.
Auto and cross power spectra of
$\gamma_{\rm obs}^i$ and $\delta_{g, \rm obs}^i$ are obtained by taking correlations, $\langle A^i B^j \rangle$,
between right hand side terms in equations~(\ref{eq:shearobs}) and
(\ref{eq:delta_g}).
Using the Limber approximation and taking upto one-loop terms, all the
relevant terms can be written in the form of,
\begin{equation}
\label{eq:Cell}
C_{AB}^{ij}(\ell)=\int~d\chi {{W_A^i(\chi) W_B^j(\chi)} \over {\chi^2}}
P_M\left(\chi,k={{\ell+1/2}\over \chi}\right),
\end{equation}
where $\chi$ is the radial comoving distance, ($A,B$) are either
($\gamma$, IA, $g$, $\mu$), $P_M$ is the matter
power spectrum of either the nonlinear model or one-loop terms, and
$W_A^i$ is the radial kernel functions:
The shear kernel is given by,
\begin{equation}
\label{eq:Wgamma}
W_{\gamma}^{i}(\chi)={{3\Omega_m }\over 2}  {H_0^2 \over c^2}
\int_\chi^\infty~d\chi_s p^i(\chi_s)
    {{D_A(0,\chi) D_A(\chi,\chi_s)} \over {a(\chi) D_A(0,\chi_s)}},   
\end{equation}
where $a(\chi)$ is the cosmic scale factor, and $D_A(\chi_1,\chi_2)$ is
the comoving angular diameter distance  between $\chi_1$ and $\chi_2$,
$p^i(\chi)$ is the redshift distribution of a tomographic galaxy sample.
The intrinsic alignment kernel is given by,
\begin{equation}
\label{eq:WIA}
W_{\rm IA}^{i}(\chi)={-A_1^i} C_1  {{\rho_{cr,0} \Omega_m} \over
    {D_+(\chi)}} p^i(\chi),
\end{equation}
where $A_1^i$ is a model parameter controlling the amplitude, $C_1$ is the
conventional constant of $C_1=5\times
10^{-14}h^{-2}M_\odot^{-1}$Mpc$^3$, $\rho_{cr,0}$ is the critical
over density at present, and $D_+(\chi)$ is the linear
growth factor normalized to unity at present.
The galaxy clustering kernel is given,
\begin{equation}
\label{eq:Wg}
W_{g}^{i}(\chi)=b_g^i p^i(\chi),
\end{equation}
and the lensing magnification kernel is, 
\begin{equation}
\label{eq:Wmu}
W_{\mu}^{i}(\chi)=2(\alpha_\mu^i-1)W_{\gamma}^{i}(\chi),
\end{equation}
where $\alpha_\mu^i$ is a local slope of the galaxy number counts $N(>f)$
at the flux limit $f_{\rm lim}$.
Introducing a shortened form in which the variable $\chi$ is
omitted from the right-hand-side of equation (\ref{eq:Cell}), the {\it
  observed} power spectra can be written by the sum of all the terms
involved:
\begin{eqnarray}
\label{eq:Cell_shear-shear}
C_{\gamma\gamma}^{ij}(\ell)&=&\int {{d\chi} \over {\chi^2}} \bigl[
  W_\gamma^i W_\gamma^j P_{\rm NL}\nonumber \\
&& +W_\gamma^i W_{\rm IA}^j P_{\rm NL}
+W_{\rm IA}^i W_{\gamma}^j P_{\rm NL}\nonumber \\
&&+W_\gamma^i b_T^j W_{\rm IA}^j P_{0|0E}
+b_T^i W_{\rm IA}^i W_\gamma^j  P_{0|0E}\nonumber \\
&&+W_{\rm IA}^i W_{\rm IA}^j P_{\rm NL}\nonumber \\
&&+W_{\rm IA}^i b_T^j W_{\rm IA}^j P_{0|0E}
+b_T^i W_{\rm IA}^i W_{\rm IA}^j  P_{0|0E}\nonumber \\
&&+b_T^i W_{\rm IA}^i b_T^j W_{\rm IA}^j P_{0E|0E}\bigr],
\end{eqnarray}
\begin{eqnarray}
\label{eq:Cell_galaxy-shear}
C_{g\gamma}^{ij}(\ell)&=&\int {{d\chi} \over {\chi^2}} \bigl[
W_g^i W_\gamma^j P_{\rm NL}+W_g^i W_{\rm IA}^jP_{\rm NL}\nonumber\\
&&+W_g^i b_T^j W_{\rm IA}^j P_{0|0E}\nonumber\\
&& +W_{\mu}^i W_{\gamma}^j P_{\rm NL} + W_\mu^i W_{\rm IA}^j P_{\rm NL}\nonumber \\
&&+W_{\mu}^i b_T^j W_{\rm IA}^j P_{0|0E}\bigr],
\end{eqnarray}
\begin{eqnarray}
\label{eq:Cell_galaxy-galaxy}
C_{gg}^{ij}(\ell)&=&\int {{d\chi} \over {\chi^2}} \bigl[ W_g^i W_g^j P_{\rm NL}
+W_g^i W_{\mu}^jP_{\rm NL}\nonumber\\
&&+W_{\mu}^i W_g^j P_{\rm NL}
+W_{\mu}^i W_{\mu}^j P_{\rm NL}\bigr],
\end{eqnarray}
where $P_{\rm NL}$ is the nonlinear model of the matter power spectrum, and
$P_{0|0E}$ and $P_{0E|0E}$ are one-loop terms defined in \citet{Blazek_2015}.
We compute the linear matter power spectrum using the publicly available
code {\tt CAMB}\footnote{{\tt https://camb.info/}} \citep{Lewis_2000}.
For the nonlinear matter power spectrum, we adopt {\tt HMcode}
\citep{Mead_2016}, implemented in {\tt CAMB} software.

%
%
\subsection{Covariance}
\label{ssec:Covariance}

The covariance matrix, which is decomposed as a sum of Gaussian, 
connected non-Gaussian and the super-sample contributions, is computed
as follows:
The Gaussian contribution is computed using {\tt NaMaster} software
\citep{Alonso_2019} with the improved narrow kernel approximation
(iNKA) estimator developed in \citet{Garc_a_Garc_a_2019,Nicola_2021}.
This estimator accounts for mode-coupling arising from the survey geometry
in a manner consistent with the pseudo-$C_{\ell}$ framework.
The connected non-Gaussian contribution and the super-sample covariance
are computed using the publicly available software {\tt OneCovariance} 
\citep{Reischke_2025}, in which the matter trispectra are derived within
the framework of the halo model.
For those two non-Gaussian terms, we only apply a scaling based on the
fraction of the observed sky, $f_{\rm sky}$.
We adopt this approximation because the non-Gaussian
contributions are relatively small compared to the Gaussian contribution
within the $\ell$-ranges under consideration (see section~\ref{ssec:datavec}).

%
%
\section{Parameter inference}
\label{sec:inference}

We employ the standard Bayesian likelihood analysis
for the cosmological inference of the set of measured 
power spectra. The log-likelihood is given by
\begin{equation}
\label{eq:lnL}
-2\ln L(p) = \sum_{i,j} (d_i-m_i(p)){\rm Cov}_{ij}^{-1} (d_j-m_j(p)),
\end{equation}
where $d_i$ is the data vector that is detailed in
Section~\ref{ssec:datavec}, 
$m_i(p)$ is the theoretical model (described in
Section~\ref{ssec:Angularpowerspecrtra}) with $p$ is a set of parameters
detailed in Section~\ref{ssec:modelparams}, and ${\rm Cov}_{ij}$ is the
covariance matrix (described in Section~\ref{ssec:Covariance}).

In order to sample the likelihood efficiently, we employ
a multimodal nested sampling algorithm
\citep{2008MNRAS.384..449F,2009MNRAS.398.1601F,2019OJAp....2E..10F}, as
implemented in the publicly available software {\tt MultiNest}.

We investigate the performance of cosmological inference using the
photo-$3\times 2$-pt, employing  mock catalogs that mimic the HSC-Y3 data in
terms of survey area and redshift distribution of galaxies.
Details regarding the construction of the mock catalogs and the results
are provided in Appendix~\ref{apdx:mock}.
In short, it is verified that unbiased estimates of
the parameters of interest can be successfully obtained through this
photo-$3\times 2$-pt cosmological inference.

%
%
\subsection{Data vector}
\label{ssec:datavec}
The data vector is constructed from the measured power spectrum after
applying scale cuts described below. 
We determined scale cuts based on two factors:
One is to avoid strongly nonlinear scales where nonlinear matter
power spectrum models have uncertainties due to baryonic feedback
effects, and the linear bias models may not be valid.
The other is to avoid potential contaminations on large scales due to
systematics, for which we use B-mode signals as an indicator.

Regarding the shear-shear power spectra, we set the scale cut to 
$317 \le \ell \le 1000$, which is covered by four band-powers.
The low-$\ell$ cut is based on the excess B-modes at scale
lower than $\ell=300$ found in \citet{Dalal_2023}, whereas the
high-$\ell$ cut is determined by considering a balance between the
uncertainties due to the baryonic feedback effect
and the number of $\ell$-modes.
Our choice of $\ell\le 1000$ is more conservative than that
of \citet{Dalal_2023} in which $\ell\le 1800$ was taken.
Although the baryon feedback effect is expected to be relatively mild in
our case, it may not be negligible \citep{Huang_2019}, thus we take
into account the baryonic feedback effect with an empirical model
(see Section~\ref{sssec:baryonparams}).

Regarding the galaxy-shear power spectra, we set the scale cut to 
$178 \le \ell \le 562$, which is covered by four band-powers.
Although the lower-$\ell$ cut is lower than that of the shear-shear
case, we confirmed that there is no evidence of excess B-mode on those
scales (see Section~\ref{ssec:bmode}).
Considering the validity of adopting a linear bias model,
we employ a more conservative high-$\ell$ cut than the shear-shear
case. 

Regarding the galaxy clustering power spectra, taking into account the
validity of the linear bias model and the fact that the Limber
approximation, which is used in calculating the theoretical models,
is not accurate for $\ell \lesssim 100$
\citep{Kilbinger_2017,Leonard_2023}, we adopt the same scale
cut as that used for the galaxy-shear power spectra,
$178 \le \ell \le 562$.
Note that unlike shear-shear and galaxy-shear cases, we do not use
cross power spectra between different tomographic bins, but we take only
the auto power spectra for the data vector.
The reason for this is as follows:
In our modeling of uncertainties in the redshift distributions of
galaxy samples, we account only for overall shifts in the
distributions  (see Section~\ref{sssec:deltazparams}).
However, while the galaxy clustering cross power spectra can be
sensitive to the tails of distributions, our modeling does not
adequately account for uncertainties in those regions.
Consequently, including these cross power spectra may introduce a bias.
Since the cross power spectra are indeed detected (see
Figure~\ref{fig:clgg_hsc}), utilizing them with more improved methods
\citep[e.g., ][]{Tanoglidis_2019,Schaan_2020,S_nchez_2023,Johnston_2025,Bernstein_2026,de_Souza_2026}
is one future direction of this study.

In summary, we have 24 power spectra, consisting of 10 shear-shear, 10
galaxy-shear, and 4 galaxy clustering, and we adopt four band powers for
each.
Therefore, our data vector consists of a total of 96 data points.

%
%
\subsection{Model parameters and their priors}
\label{ssec:modelparams}

In this subsection, we describe model parameters and their prior
distributions used in our cosmological analysis.
These are summarized in Table~\ref{table:parameters}.

%
\begin{table}
  \caption{Summary of model parameters and their prior ranges.
    \label{table:parameters}} 
\begin{tabular}{ll}
\hline
Parameter & Prior\\
\hline
\multicolumn{2}{l}{\bf Cosmological parameters (Section~\ref{sssec:cosmoparams})}\\
$\Omega_c$ & flat($0.1,0.7$) \\
$\omega_b=\Omega_b h^2$ & flat($0.02,0.025$) \\
$h$ & flat($0.62,0.80$) \\
$A_s\times 10^{-9}$ & flat($0.5,10$) \\
$n_s$ & flat($0.87,1.07$) \\
\hline
\multicolumn{2}{l}{\bf Intrinsic alignment parameters
  (Section~\ref{sssec:IAparams})}\\
$A_1^i$ & flat($-3,3$) \\
$b_T^i$ & flat($0,2$) \\
\hline
\multicolumn{2}{l}{\bf Galaxy clustering bias
  (Section~\ref{sssec:galaxybiasparams})}\\
$b_g^i$ & flat($0.5,3.5$) \\
\hline
\multicolumn{2}{l}{\bf Magnification effect parameters
  (Section~\ref{sssec:magnifparams})}\\
$\alpha_\mu^i$& flat($0,2$) \\
\hline
\multicolumn{2}{l}{\bf Baryonic feedback model parameter
  (Section~\ref{sssec:baryonparams})}\\
$A_{\rm bary}$ & flat($2,3.13$) \\
\hline
\multicolumn{2}{l}{\bf Redshift distribution shift parameters
  (Section~\ref{sssec:deltazparams})}\\
$\Delta z_1$& Gauss($0,0.024$) \\
$\Delta z_2$& Gauss($0,0.022$) \\
$\Delta z_3$& flat($-0.5.0.5$) \\
$\Delta z_4$& flat($-0.5.0.5$) \\
\hline
\end{tabular}
\end{table}

%
%
\subsubsection{Cosmological parameters}
\label{sssec:cosmoparams}

We focus on the flat $\Lambda$CDM cosmological model characterized
by five parameters;
the density parameter of CDM ($\Omega_c$),
the normalization of matter fluctuation $A_s$ (the scalar amplitude of
the linear matter power spectrum at $k=0.05$~Mpc$^{-1}$), 
the density parameter of baryons ($\Omega_b$),
the Hubble parameter ($h$), 
and the scalar spectrum index ($n_s$).
Among those parameters, the photo-$3\times 2$-pt is most sensitive to
$\Omega_c$ and $A_s$, or the derived parameter $\sigma_8$.
Thus we adopt prior ranges that are sufficiently wide for these
parameters (see Table~\ref{table:parameters}).
For ($\Omega_b$, $n_s$, and $h$), which are only weakly constrained with
photo-$3\times 2$-pt, we set prior ranges which largely bracket
allowed values from external experiments (see
Table~\ref{table:parameters}).
For the sum of neutrino mass, we take a fixed value of $\sum m_\nu
=0.06$~eV from the lower bound indicated by
the neutrino oscillation experiments \citep[e.g.,][for a
  review]{2013neco.book.....L}.

%
%
\subsubsection{Intrinsic alignment model parameters}
\label{sssec:IAparams}

We adopt the tidal alignment model, a linear intrinsic alignment model
that incorporates the non-linear matter power spectrum
\citep{Hirata_2004,2007NJPh....9..444B,2015SSRv..193....1J}, extended to
the one-loop level \citep{Blazek_2015}.
The relevant expressions are summarized in
Section~\ref{ssec:Angularpowerspecrtra}.
This model has two parameters, which we treat independently for each
tomographic sample, denoted by
$A_1^i$ and $b_T^i$.
For $A_1^i$, a flat prior ranging from $-3$ to 3 is taken to
cover a sufficiently wide parameter range.
For $b_T^i$, a flat prior ranging from 0 to 2 is taken to
cover a physically reasonable parameter range.
Indeed, within this prior range, our data vector does not have
practical sensitivity to this parameter.
Therefore, it has no practical influence on constraints on parameters
that we are mainly interested in.
We check this by performing test runs with fixed values of
$b_T^i=0$ and $b_T^i=1$
(see Section~\ref{ssec:SystematicsTests}).

%
%
\subsubsection{Galaxy clustering bias parameters}
\label{sssec:galaxybiasparams}

We adopt the scale independent linear bias model for galaxy clustering.
The bias parameter, $b_g^i$, is treated independently for each
tomographic sample.
We take a flat prior ranging from $0.5$ to 3.5 to cover a sufficiently
wide parameter range.

%
%
\subsubsection{Lensing magnification effect parameters}
\label{sssec:magnifparams}

%
%
\begin{figure}
\begin{center}
\includegraphics[width=82mm]{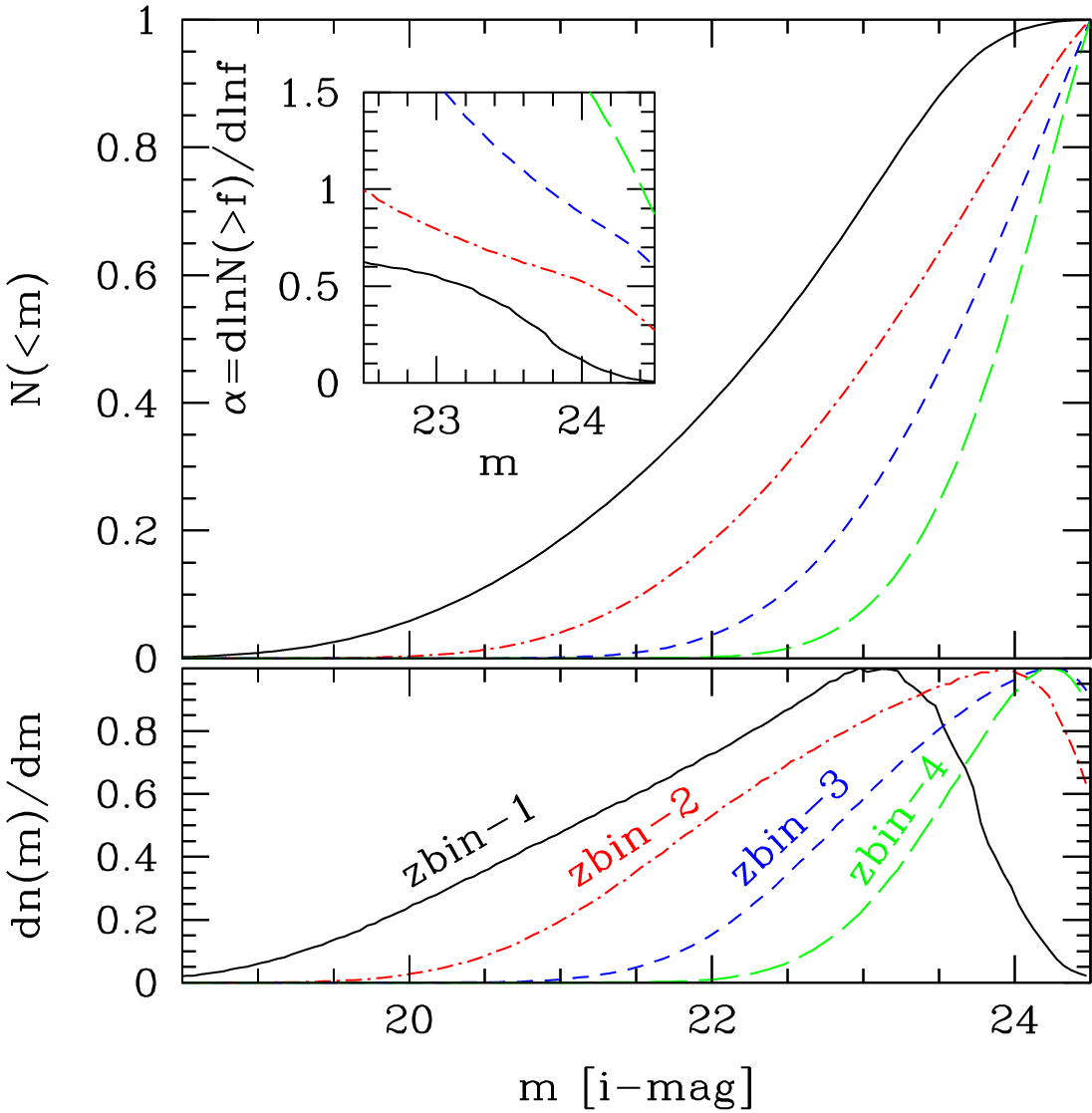}
\end{center}
\caption{Normalized number counts of galaxies in four tomographic
  samples as a function of $i$-band magnitude. Top and bottom panels
  show accumulated and differential counts, respectively.
  Inserted plot shows the slope of the counts in flux 
  $\alpha_\mu^i=d\ln N(>f) /d\ln f$, a key factor determining the
  strength of the gravitational lensing magnification effect (see
  section \ref{sssec:magnifparams}).   
  \label{fig:alpha_mu}}
\end{figure}

The lensing magnification effect alters the galaxy density fluctuation
field depending on a local slope of the galaxy number counts $N(>f)$
at the flux limit $f_{\rm lim}$, denoted by
$\alpha_\mu^i= d\ln N(>f) /d\ln f$ \citep{2001PhR...340..291B}. 
Therefore the size of the effect varies depending on the sample.
Since the number counts $N(>f)$ in the above expression should be {\it
  unlensed} one, an exact value can not be estimated by {\it
  observed} counts, but may give an approximate estimate.
We derived the slopes using {\it observed} counts, and found;
$\alpha_\mu^i \sim 0$, 0.3, 0.6, and 0.9 for $z$-bin $i=1$, 2, 3, and 4,
respectively (see Figure~\ref{fig:alpha_mu}).
Considering these results, we treat $\alpha_\mu^i$ independently for
each tomographic sample, and take a flat prior ranging from 0 to 2.
Note that declines seen in the differential number counts at faint
magnitudes are due to the resolution factor cut imposed in making
HSC-Y3 shape catalog \citep{Li_2022}.
The resolution factor is a ratio between the size of PSF and the size of
galaxy, and is used to quantify the extent to which the galaxy is
resolved compared to the PSF.
Due to this cut, smaller (and thus generally fainter) galaxies compared
with the PSF size were removed from the catalog.
As a result, the galaxy number counts decreases at fainter magnitudes.
This cut has a strong impact especially on low-$z$ samples, leading to a
turnover at a brighter magnitude. 

%
%
\subsubsection{Baryonic feedback model parameter}
\label{sssec:baryonparams}

In modeling the baryonic effect, we follow the methodology of
\citet{2015MNRAS.454.1958M}, in which a modification 
of the dark matter power spectrum due to supernovae and active galactic
nuclei is modeled by the fitting function based on the halo model with
the halo bloating parameter ($\eta_b$) and the amplitude of the halo
mass concentration ($A_{\rm bary}$).
Following \citet{2021A&A...646A.129J}, we adopt the empirical
relation, 
\begin{equation}
\label{eq:HMbaryon}
\eta_b = 0.98-0.12 A_{\rm bary}.
\end{equation}
The value of $A_{\rm bary}=3.13$ corresponds to an absence of baryonic
feedback. 
We take a flat prior ranging from $2$ to 3.13 following HSC-Y3 cosmic
shear analyses \citep{Dalal_2023,Li_2023}.

%
%
\subsection{Systematics parameters}
\label{ssec:systematics}

This subsection summarizes our treatment of systematic uncertainties
originating from source galaxy redshift distributions,  
PSF leakage and PSF modeling errors, and shear multiplicative bias
corrections. 

%
%
\subsubsection{Redshift distribution uncertainties}
\label{sssec:deltazparams}

Regarding uncertainties in redshift distributions
of source galaxies, we follow the approach used in the HSC-Y3 cosmic
shear analyses \citep{Dalal_2023,Li_2023}.
Specifically, we introduce a nuisance parameter, $\Delta z_i$, for each
tomographic redshift bin to shift the source redshift distribution such
that $p^i(z) \rightarrow p^i(z + \Delta z_i)$.
Following \citet{Dalal_2023,Li_2023}, we adopt Gaussian priors centered
at zero with widths of $\sigma = 0.024$ and $0.022$ for $\Delta z_1$ and
$\Delta z_2$, respectively,
For the higher-redshift bins, $\Delta z_3$ and $\Delta z_4$, we adopt
flat priors ranging from $-0.5$ to 0.5.
The reason for this choice is as follows:
As described in \citet{Dalal_2023,Li_2023}, the redshift distributions
were obtained from a joint estimation using photo-z and
cross-correlation between the HSC-Y3 shape catalog and the CAMIRA-LRG
catalog which covers the redshift upto 1.2 \citep[see for
  details][]{Rau_2023}, thus $p^i(z)$ of the 3rd and 4th redshift bin
were not fully calibrated, and may have potential biases.
In order to avoid these biases, wide flat priors were adopted for 
$\Delta z_3$ and $\Delta z_4$.
Indeed, \citet{Dalal_2023} and \citet{Li_2023} found their mean 1D
posteriors to be $\Delta z_3\sim-0.07$ and $\Delta z_4\sim -0.16$.
We take the above mentioned prior range to well include these results.

For the redshift distributions of galaxy clustering samples used to  
compute the galaxy clustering and magnification kernels (equations
(\ref{eq:Wg}) and (\ref{eq:Wmu}), respectively), 
we adopt the same distributions as those used for the shape galaxy samples.
It should be noted that although we use the same galaxy
samples for both the shear and galaxy clustering measurements, their
redshift distributions may differ due to two factors:
First, area cuts are applied to the galaxy clustering samples;
second, the shape weights are applied to the shear measurement, but not
to the clustering measurements.
We check these points by comparing the stacked redshift probability
distribution functions (PDFs) of galaxies in each sample.
These stacked redshift PDFs are computed using the individual redshift
probabilities estimated by the {\tt dNNz} photo-$z$ method, which are
included in the HSC-Y3 products.
The stacked redshift PDFs for the shear samples are estimated by
accounting for the shape weights, whereas those for the galaxy clustering
samples are estimated using galaxies survived after the area cuts 
without accounting for the shape weights.
It is found that differences between the two PDFs are very small;
their overall shapes are very similar, though the high-$z$ tails of the
shear samples are slightly suppressed.
Quantitatively, the differences between the mean redshifts of the two
PDFs are
$|\Delta z_{\rm mean}^i|<6\times 10^{-3}$ or
$|\Delta z_{\rm mean}^i/z_{\rm mean}^i|<6\times 10^{-3}$.
Based on these results, we conclude that neither the area cuts nor
the shape weight significantly alter the redshift distributions of galaxy
samples.
Note that it is found that the impact of the area cuts is negligible;
when considering only the area cuts, $|\Delta z_{\rm mean}^i|<7\times 10^{-4}$ or
$|\Delta z_{\rm mean}^i/z_{\rm mean}^i|<9\times 10^{-4}$.

In summary, we adopt the same redshift distributions with common shift
parameters for both the shape and clustering tomographic
galaxy samples.
Specifically, when computing the four kernel functions,
equations (\ref{eq:Wgamma})-(\ref{eq:Wmu}),
the identical $p^i(z)$ are used with the common shift parameters,
$\Delta z_i$, applied consistently to all of them.

%
%
\subsubsection{Corrections for PSF related errors and shear multiplicative bias}
\label{sssec:PSFandMbias}

Regarding systematic effects arising
from uncertainties in the PSF induced additive biases, we use
the result of \citet{Dalal_2023}.
As described in Section~IV-D of \citet{Dalal_2023}, they modeled
PSF-related errors by using PSF auto, and PSF-galaxy correlations with
nuisance parameters, which were marginalized over after Bayesian
parameter sampling. 
Since the impact of the PSF-related errors on constraints on parameters
of interest is found to be small (see Figure~11 and Table~IV of
\citet{Dalal_2023}, and also see section \ref{sec:dataset} for a brief
summary ), we take fixed values derived in
\citet{Dalal_2023};  
to be specific, we adopt the best fit additive biases from the
PSF-related systematics, $\Delta C_\ell$, shown in Figure~3 of
\citet{Dalal_2023}.

Regarding systematic effects arising from the uncertainty in
the shear multiplicative bias, we again adopt
the result of \citet{Dalal_2023}.
As described in Section~V-F of \citet{Dalal_2023}, they modeled
the shear multiplicative bias by introducing nuisance parameters, which
were marginalized over after Bayesian parameter sampling.
Since its impact on parameter constraints is small, we take fixed
values for multiplicative biases taken from posterior modes reported in
Table~VI of \citet{Dalal_2023}.
Note that in \citet{Dalal_2023}, the multiplicative biases are treated
as nuisance parameters with Gaussian prior with the mean
of 0 and $\sigma = 0.01$. Therefore the multiplicative bias correction
introduces additional errors in parameter constraints, which are not
included in our analysis, leading to underestimation of errors.
However since the prior width of $\sigma = 0.01$ is much smaller than
the size of our covariance matrix, the size of the underestimation is
small.

%
%
\section{Results}
\label{sec:results}

%
%
\subsection{Parameter constraining power of galaxy-shear and galaxy
  clustering power spectra}
\label{ssec:constraingpower}

Before presenting the results, it is instructive to provide an overview of
the enhanced parameter constraining power achieved by adding
galaxy-shear and galaxy clustering power spectra to the cosmic
shear-only analysis
\citep[see][for previous studies examining similar methods]{Schaan_2020,Johnston_2025}.
In short, the added power spectra enhance the constraining power on
particular nuisance parameters that increase the error on $S_8$ in
cosmic shear-only analyses.
Consequently, constraints on these nuisance parameters become tighter, leading
to an improved $S_8$ constraint. 
It is the intrinsic alignment parameter ($A_1^i$) and
the shift parameter of the redshift distributions of galaxy samples
($\Delta z_i$) that the added power spectra can particularly tighten
their constraints.
We explain it in more detail below.

It is $S_8$ parameter among other cosmological parameters that cosmic shear
analyses have the highest sensitivity, but uncertainties due to the
intrinsic alignments increase an error on $S_8$ without introducing
significant bias.
The galaxy-shear auto-power spectra provide useful information on the
intrinsic alignment as their signals mostly come from both
the galaxy-shear and galaxy-IA correlations.
In our case, the galaxy-shear contribution is larger than the galaxy-IA,
but the size of galaxy-IA contribution is about $10-50$ percent of that of
galaxy-shear (see Figure~\ref{fig:clge_hsc}), whereas in the cosmic
shear power spectra, the shear-IA contribution is $\lesssim 30$ percent
of that of shear-shear.
Therefore, adding the galaxy-shear power spectra can reduce errors on IA
parameters, resulting in an improved constraint on $S_8$ compared with
cosmic shear-only analyses. 

The uncertainties in the redshift distributions of source galaxies can
be a serious cause of a bias in $S_8$ estimate.
In HSC-Y3 cosmic shear analyses \citep{Dalal_2023,Li_2023}, a
conservative wide prior range for the shift parameters $\Delta z_3$ and
$\Delta z_4$  
was taken to avoid a bias, which however leads to $\sim 25$ percent
enlargement of $S_8$ error compared with the case with informative
$\Delta z_i$ priors for those bins.
The galaxy clustering auto-power spectra (and to a lesser extent, 
the galaxy-shear power spectra) have a good sensitivity to
$\Delta z_i$, because
the galaxy clustering kernel (equation~(\ref{eq:Wg})) is directly
related to the redshift distributions, $p(z)^i$.
Therefore, in photo-$3\times 2$-pt analyses, the galaxy clustering power
spectra serve as a ``self-calibration'' for $\Delta z_i$ parameters,
which can reduce errors on $\Delta z_i$, resulting in an improved
constraint on $S_8$. 
\citep[see][for previous studies exploring the 
    self-calibration of photometric galaxy redshift distributions
    using similar methods]{Schaan_2020,Johnston_2025}.

It should be noted that the galaxy clustering involves the
galaxy bias parameter $b_g^i$, and the additional contribution from
the lensing magnification effects (see equation~(\ref{eq:delta_g})),
which introduce additional degrees of freedom in parameter inference.
Therefore, it is not obvious if combining galaxy-shear and galaxy
clustering power spectra with a cosmic shear analysis can enhance its 
parameter constraining power.
In our case, qualitatively speaking, the increase in degrees of freedom
is counterbalanced by the increased amount of information, resulting
in no degradation of the original parameter constraining power.

%
%
\begin{figure}
\begin{center}
\includegraphics[height=82mm,angle=270]{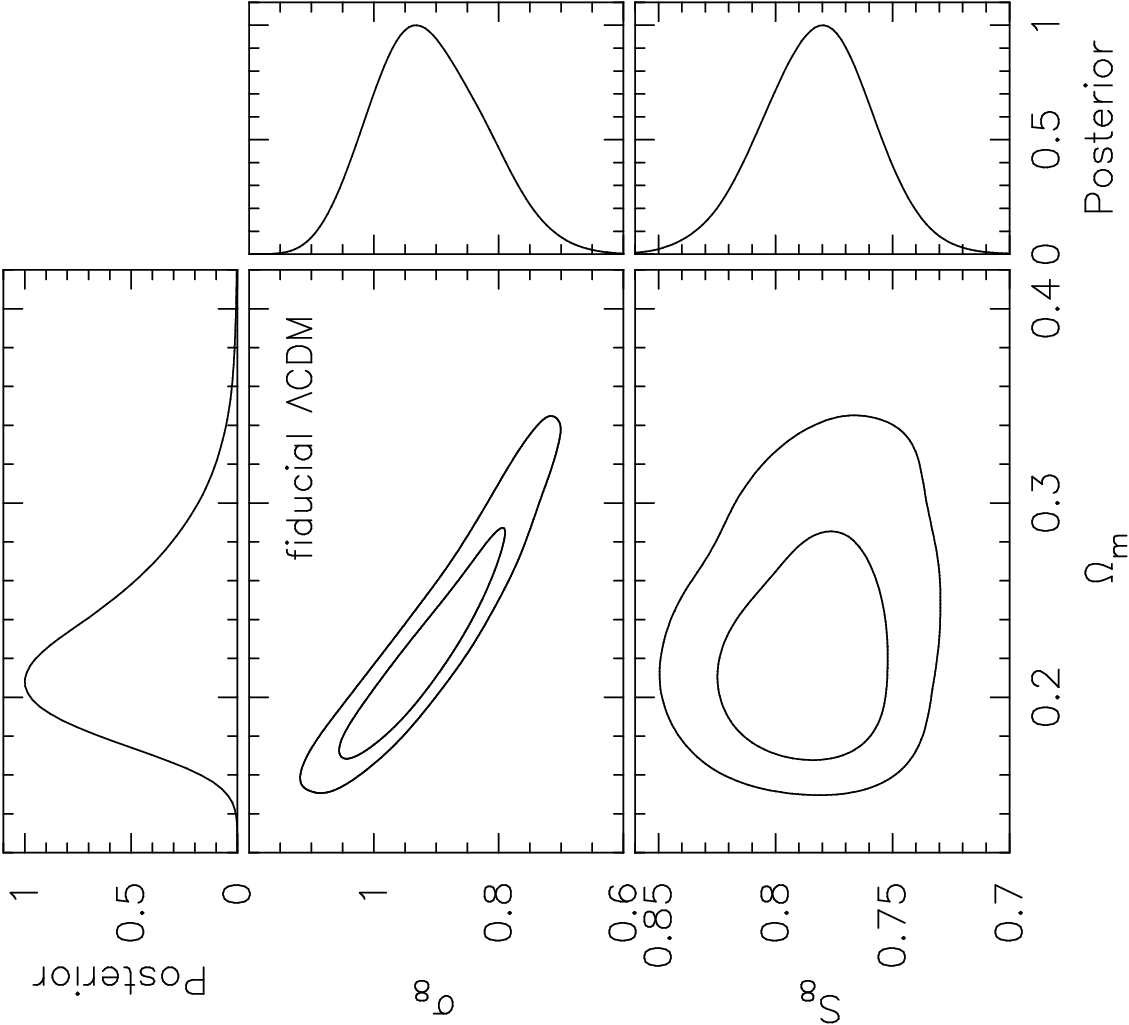}
\end{center}
\caption{Marginalized two-dimensional posterior contours (68\% and 95\%
  credible levels) for the fiducial $\Lambda$CDM model, shown in the
  $\Omega_m$-$\sigma_8$ plane (middle left) and the $\Omega_m$-$S_8$
  plane (bottom left). 
  The corresponding marginalized 1D posterior distributions are shown
  alongside the 2D plots.  
  \label{fig:om_s8_sig8}}
\end{figure}

%
%
\begin{figure*}
  \begin{center}
    \includegraphics[height=140mm,angle=270]{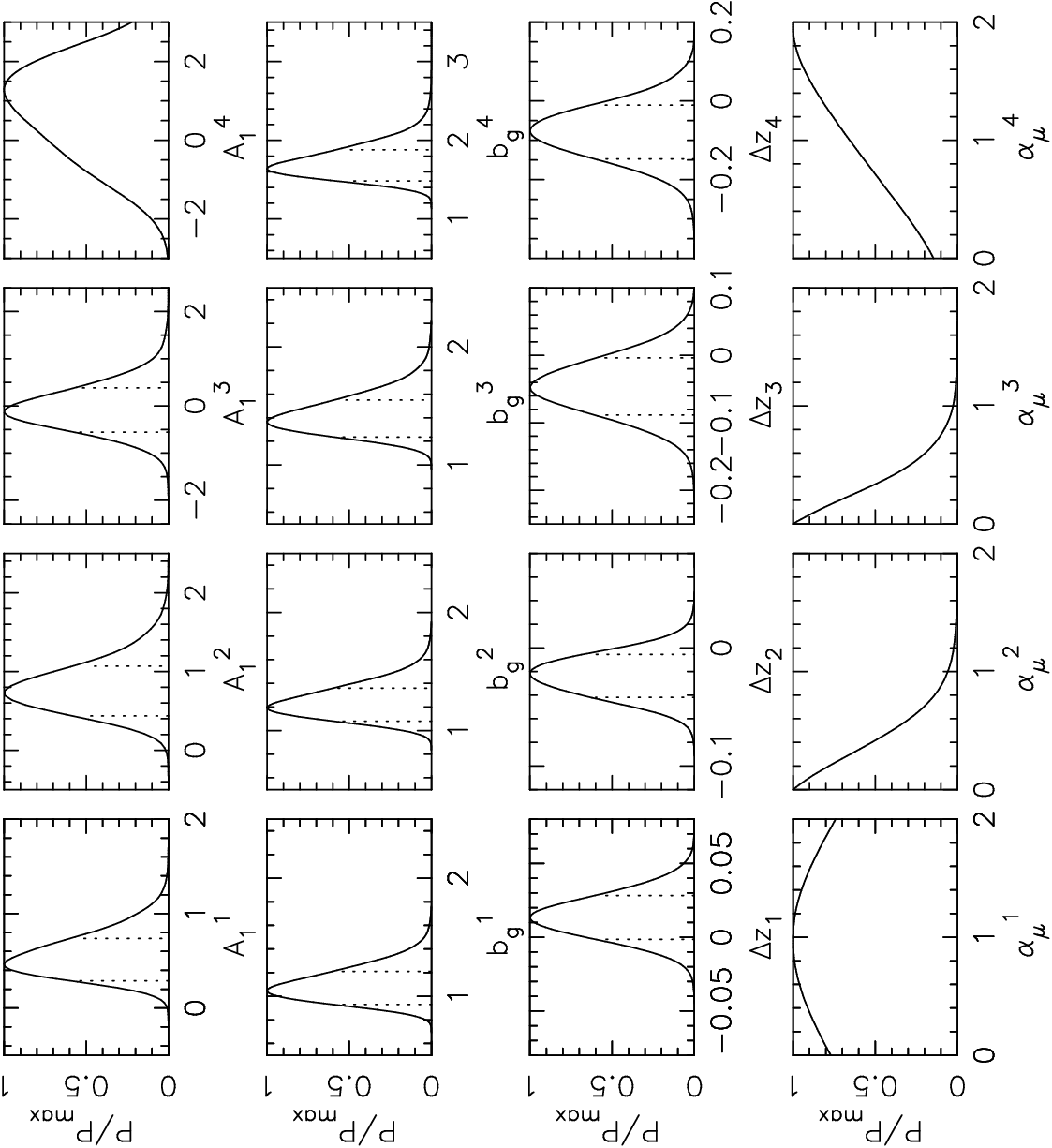}
  \end{center}
  \caption{The marginalized one-dimensional
    posterior distributions for the fiducial $\Lambda$CDM model.
    From the top to bottom rows, the panels display the intrinsic
    alignment amplitude parameters,
    the galaxy clustering bias parameters,
    the shift parameters of the galaxy redshift distributions,
    and the lensing magnification effect parameters. 
    Dotted vertical lines represent the approximate 68\% credible
    intervals, which are omitted for poorly constrained parameters.    
    \label{fig:post_4x4}}
\end{figure*}

%
%
\subsection{Parameter constraints in the fiducial $\Lambda$CDM model}
\label{ssec:ParameterConstraints}

Here we present results for our fiducial $\Lambda$CDM model.
We evaluate the goodness-of-fit with the standard $\chi^2$ test;
the $\chi^2$ value for the best-fitting
parameter set is $\chi^2 = 70.5$ for the {\it effective} degree of
freedom\footnote{Although the total number of model parameters is 26 for
our fiducial case, only 11 of them ($\Omega_c$, $A_s$, $A_1^{1-3}$,
$b_g^{1-4}$, and $\Delta z_{3-4}$) are constrained by the data with
much narrower posterior distributions than priors. Therefore, a
conservative choice of the effective number of free parameters 
should account for only these 11 parameters.} of $96-11=85$,
resulting in a $p$-value of 0.87. Therefore we conclude that the model
provides a good fit to the data.

Figure~\ref{fig:om_s8_sig8} shows marginalized posterior contours in the
$\Omega_m$-$\sigma_8$ and $\Omega_m$-$S_8$ planes, along with the
marginalized one-dimensional posterior distributions for each parameter.
We find marginalized 68\% confidence intervals of
$0.18<\Omega_m<0.25$, $0.83<\sigma_8< 1.01$, and
$0.76 < S_8 <0.81$.
We will discuss the robustness of the result against various
systematics in modeling in section \ref{ssec:SystematicsTests}, and
will compare with HSC-Y3 cosmic shear results in section
\ref{ssec:Comparison}. 

Marginalized one-dimensional posterior distributions of 16 model
parameters ($A_1^i$, $b_g^i$, $\Delta z_i$, and $\alpha_\mu^i$) are
shown in Figure~\ref{fig:post_4x4}. 
In those plots, dotted vertical lines represent the approximate 68\%
credible intervals, which are not shown for poorly constrained
parameters.
It is found from the plots that the IA parameter, $A_1^i$, for
the highest redshift bin and the lensing magnification
parameter $\alpha_\mu^i$ for all the bins are poorly constrained.
Regarding the IA parameter, the constraints on $A_1^i$ become weaker for
higher redshift bins.
There are two reasons for this:
The first concerns the cosmic shear $C_\ell$;
the contribution from shear-IA becomes smaller relative to the
contribution from shear-shear for higher redshift bins.
The second concerns the galaxy-shear $C_\ell$; the errors on $C_\ell$s
become larger for higher redshift bins as shown in Figure~\ref{fig:clge_hsc}. 
Both the two factors lead to weaker constraints on $A_1^i$ in the
higher-redshift bins.
Regarding the lensing magnification parameter, the constraints on
$\alpha_\mu^i$ for all the bins are very week.
This is because the lensing magnification effect is
small for galaxy clustering auto spectra as demonstrated in
Figure~\ref{fig:clgg_hsc}. 
Although constraints on $\alpha_\mu^i$ are very weak,
no discrepancy is observed between them and the values estimated from
galaxy number counts (see section~\ref{sssec:magnifparams} and
Figure~\ref{fig:alpha_mu}). 

%
%
\begin{figure}
\begin{center}
\includegraphics[height=82mm,angle=270]{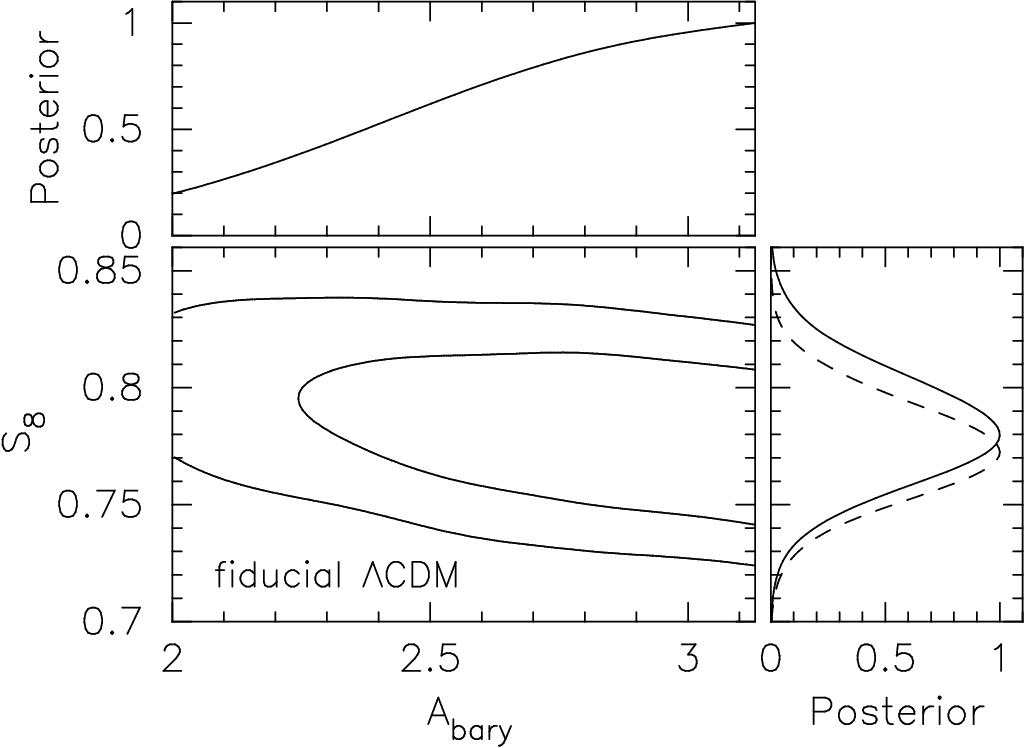}
\end{center}
\caption{Marginalized two-dimensional posterior contours (68\% and 95\%
  credible levels) in the $A_{\rm bary}$-$S_8$ plane for the fiducial
  $\Lambda$CDM model are shown.
  The corresponding marginalized 1D posterior distributions are shown
  alongside the 2D plots.
  The dashed line in the right panel is for the model without the
  baryonic feedback effect ($A_{\rm bary}$ is fixed at 3.13).  
  \label{fig:abar_s8}}
\end{figure}

Figure~\ref{fig:abar_s8} shows the marginalized posterior contours in the
$A_{\rm bary}$-$S_8$ plane, along with the corresponding marginalized
one-dimensional posterior distributions.
The figure confirms the expected negative correlation between
$A_{\rm bary}$ and $S_8$; however, the correlation is very weak, and no
useful constraint on $A_{\rm bary}$ is obtained.
We discuss the impact of uncertainties in the baryonic feedback
effect on our results in the next subsection.

%
%
\subsection{Systematics tests}
\label{ssec:SystematicsTests}

\begin{figure}
\begin{center}
\includegraphics[height=82mm,angle=270]{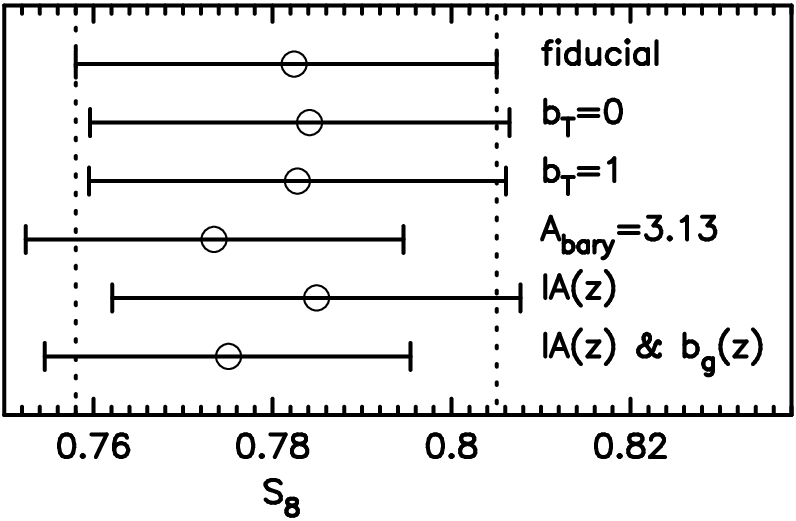}
\end{center}
\caption{Means and 68\% credible intervals of marginalized
  one-dimensional constraints on $S_8$. The fiducial case (top) is
  compared with different setups to check the robustness of the result.
  Vertical dotted lines show the 68\% credible interval of the fiducial
  case.  
\label{fig:s8ranges_sys}}
\end{figure}

In this subsection, we evaluate how uncertainties in our theoretical
models impact cosmological inference.
Specifically, we investigate the effects of the IA bias parameter $b_T^i$, the
baryonic feedback parameter $A_{\rm bary}$, and our modeling
choices for both IA and galaxy clustering bias.
To assess the impact of uncertainties in
our models, we focus on the $S_8$ constraint as it is a primary parameter
to be constrained by photo-$3\times 2$-pt.
The results of these systematics tests
are summarized in Figure~\ref{fig:s8ranges_sys}, where 68\% credible
intervals of $S_8$ derived from systematics tests are compared with the
fiducial result.
In short, our $S_8$ constraint is robust against the
systematic uncertainties considered, as described in detail below. 

%
%

For the IA model, we adopt the NLA model including higher order terms
(described in section~\ref{ssec:Angularpowerspecrtra}).
This model includes the bias parameter $b_T^i$, representing the biased
relation between the galaxy number density field and IA strength, which
arises from the galaxy density-weighted IA
(see section \ref{ssec:Angularpowerspecrtra}, for details). 
The prior range of $b_T^i$ is, however, not well understood.
We adopt a flat prior of $0\le b_T \le 2$.
The resulting posterior distributions for $b_T^i$ are very broad and 
are truncated at the boundaries of the prior range.
Furthermore, it is found that $b_T^i$ correlates only very weakly with $S_8$.
Those results can be understood as follows:
All terms involving $b_T^i$ are higher order terms (see
equations~(\ref{eq:Cell_shear-shear}) and (\ref{eq:Cell_galaxy-shear})),
and their contribution to the total power spectra within the
$\ell$-ranges of the data vector is minor.
Consequently, variations in $b_T^i$ have almost no impact on the
results.
To check this interpretation, we perform two cosmological inferences with  
$b_T^i$ being fixed at either 0 or 1.
The results are shown in Figure~\ref{fig:s8ranges_sys}, where 68\%
credible intervals on $S_8$ obtained from those test cases are compared
with the fiducial result.
We find that in both the cases, changes in $S_8$ constraint are very
small, supporting our interpretation.

For the model of the baryonic feedback effect, we adopt the empirical
approach based on the halo model
\citep[][see section \ref{sssec:baryonparams} for details]{2015MNRAS.454.1958M,2021A&A...646A.129J}, which incorporates a single parameter, $A_{\rm bary}$.
We adopt a flat prior of $2\le A_{\rm bary} \le 3.13$ following HSC-Y3 cosmic
shear analyses \citep{Dalal_2023,Li_2023}.
To assess the impact of baryonic feedback effect, we perform a 
cosmological inference without the baryonic effect, that is $A_{\rm bary}$
being fixed at $3.13$, which can be regarded as an extreme case.
The result is shown in Figures~\ref{fig:s8ranges_sys} and
\ref{fig:abar_s8}.
As expected, ignoring baryonic effects shifts the $S_8$ constraint
toward smaller values.
The shift of the mean of $S_8$ is not significant and is $0.38 \sigma$.
Furthermore, as can be seen from Figure~\ref{fig:abar_s8},
the 1D posterior distribution of $A_{\rm bary}$ does not show a particular
preference for smaller values of $A_{\rm bary}$ (i.e., stronger baryonic
effect), though the posterior distribution is broad and truncated
at the lower bound.
Based on those results, we conclude that the effect of baryonic feedback
on our fiducial cosmological constraints is not significant.

Next, we investigate the effects of our modeling of IA and galaxy
clustering bias on the cosmological inference.
For these models, we employed parameters independent for each redshift
bin, instead of a redshift power-law model, i.e, $p\propto (1+z)^q$.
While this model allows for a high degree of freedom\footnote{It
  should be noted that the constraints on $b_g^i$ are 
  strongly correlated with each other through the strong correlations between
  all four $b_g^i$ and $\sigma_8$, whereas those on
  $A_1^i$ are not.}, it may carry
the risk of excessive, non-physical degrees of freedom.
We check this point by performing cosmological inferences with  
redshift power-law models.
First, we switch the IA model to the redshift power-law model, which
is widely adopted in cosmic shear studies,
\begin{equation}
  \label{eq:iaz}
  A_1(z)=A_1 \left({{1+z} \over {1+z_0}}\right)^{\eta_1}, 
\end{equation}
where $z_0=0.62$, and another parameter, $b_T$, is treat as a constant
common to all the four redshift bins.
The result is shown in Figure~\ref{fig:s8ranges_sys}
with the label ``IA$(z)$''.
It is found from the figure that the constraint on $S_8$ is consistent
with the fiducial model, thus we may conclude that the fiducial model does not
yield biased results. 
The resulting 68\% credible intervals of model parameters are;
$0.33 < A_1 <  0.77$, $-1.25 < \eta_1 <  1.81$, and $b_T < 0.98$ (the upper
bound only, the lower side of the posterior is truncated by the prior of
$0 \le b_T \le 2$).
The poor constraint on $\eta_1$ is expected as the constraints on
$A_1^i$ for two higher redshift bins are poor.

In addition to the IA model, we switch the galaxy bias model to the
redshift power-law model,
\begin{equation}
  \label{eq:bgz}
  b_b(z)=b_g (1+z)^\beta.
\end{equation}
The result is shown in Figure~\ref{fig:s8ranges_sys}
with the label ``IA$(z)$ \& $b_g(z)$'', in which it is found that the
constraint on $S_8$ is consistent  with the fiducial model.
The resulting 68\% credible intervals of model parameters are;
$0.54< b_g < 0.68$, and $0.89 < \beta < 1.04$.
Therefore, it is found that this model yields bias values consistent
with those obtained from the fiducial model.
These results indicate that the redshift power-law model provides a
reasonable description of the galaxy clustering bias.

%
%
\subsection{Comparison to the HSC-Y3 cosmic shear results}
\label{ssec:Comparison}

%
%
\begin{figure}
\begin{center}
\includegraphics[height=82mm,angle=270]{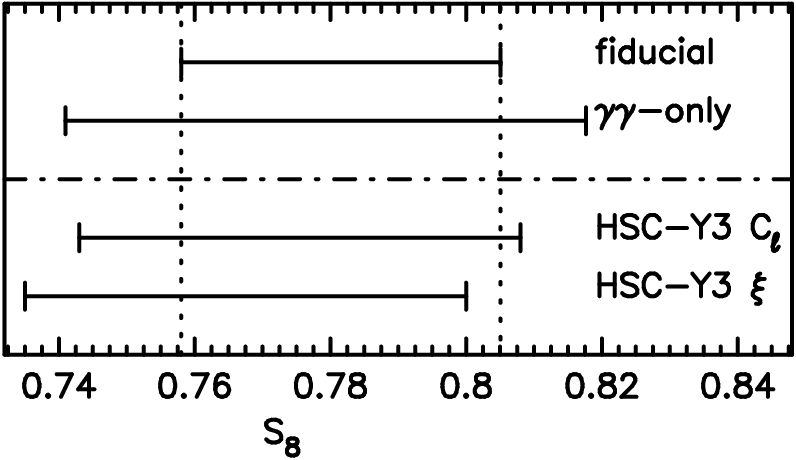}
\end{center}
\caption{Same as Figure \ref{fig:s8ranges_sys} but our fiducial $S_8$
  constraint is compared with results from the HSC-Y3
  cosmic shear studies \citep[][the bottom two
    rows]{Dalal_2023,Li_2023}.
  Note that central values are not shown to avoid possible
  misunderstanding, as those studies adopt mode of the marginalized
  posterior distribution for the central value, instead of mean
  adopted in this study.
  The case labeled ``$\gamma\gamma$-only'' is the result from our own
  cosmic shear-only analysis adopting our fiducial
  parameter setup and scale cut ($317 \le \ell \le 1000$).  
  \label{fig:s8ranges_hscy3}}
\end{figure}

Finally, we compare our $S_8$ constraint with those from the HSC-Y3
cosmic shear studies \citep{Dalal_2023,Li_2023},
and with the result from our own cosmic shear-only analysis adopting
our fiducial parameter setup and scale cuts ($317 \le \ell \le 1000$).
Results, shown in Figure~\ref{fig:s8ranges_hscy3}, are very
reasonable:
The $S_8$ constraint from our cosmic shear-only analysis is looser than
those from the HSC-Y3 cosmic shear studies as our analysis use narrower
$\ell$-range than theirs.
Nevertheless, the $S_8$ constraint from our fiducial case is $\sim 25$\%
tighter than the HSC-Y3 cosmic shear results, 
demonstrating the effectiveness of the photo-$3\times 2$-pt.
It should be noted that errors in this study could be slightly underestimated 
because compared to the HSC-Y3 cosmic shear studies, our treatment of
the PSF systematic and shear calibration biases is simplified.
Nevertheless, the effectiveness of the photo-$3\times 2$-pt is 
demonstrated through an internal comparison between the fiducial result and the
result from our cosmic shear-only analysis. 

Another important point in the comparison with the HSC-Y3 cosmic shear
studies is the constraints on $\Delta z_3$ and $\Delta z_4$ (shown in
Figure~\ref{fig:post_4x4}).
Our constraints on those parameters are
consistent with the results of the HSC-Y3 cosmic shear studies, but our
credible intervals are $\sim 30$\% tighter.
This improved constraining power on the shift parameters is an advantage
of the photo-$3\times 2$-pt over cosmic shear-only analysis.

%
%
\section{Summary and discussion}
\label{sec:summary}

We have presented a cosmological analysis of the photo-$3\times 2$-pt
measured from the HSC-Y3 data, covering 433 deg$^2$.
Photometric redshifts derived from the HSC
five-band photometry are adopted to select galaxies and
divided into four tomographic redshift bins ranging from
$z = 0.3$ to $1.5$ with equal widths of $\Delta z = 0.3$.
The total number of selected galaxies is 25 million.

We measure three types of angular power spectra; the galaxy clustering,  
galaxy-shear (galaxy-galaxy lensing), and shear-shear (cosmic shear).
They have been computed using the pseudo-$C_\ell$ formalism implemented
in {\tt NaMaster} software 
\citep{Alonso_2019,Nicola_2020,Nicola_2021}.
In deriving the galaxy clustering power spectra, we mitigate effects of
systematics in two steps, one is to cut areas 
most likely affected by systematics, and the other is to deproject the
maps of these systematics from the galaxy density maps using ``template
deprojection'' method developed in \citet{Elsner_2016}, and implemented
in NaMaster software.

In determining scale cuts of power spectra, we take into account two
points: One is to avoid strongly nonlinear scales where nonlinear matter
power spectrum models have uncertainties due to baryonic feedback
effects, and the linear bias model may not be valid.
The other is to avoid potential contaminations on large scales due to
systematics, for which we use B-mode signals as an indicator.
We have set the scale cut to $317 \le \ell \le 1000$ for cosmic shear,
and to $178 \le \ell \le 562$ for galaxy clustering and galaxy-shear power
spectra.
We evaluated the consistency of the BB- and EB-mode cosmic shear power
spectra, as well as the B-mode galaxy-shear power spectra, with zero
using standard $\chi^2$ statistics.
In all cases, we found no evidence of a significant B-mode signal.

We have performed a standard Bayesian likelihood
analysis for the cosmological inference of the measured
photo-$3\times 2$-pt data.
Our fiducial $\Lambda$CDM model consists
of five cosmological parameters and 21 nuisance parameters including 
the intrinsic alignment model parameters, galaxy clustering biases, lensing
magnification effect parameters, shift parameters of a galaxy redshift
distributions, and baryonic feedback effect parameter.
We have found that our fiducial model
fits the measured photo-$3\times 2$-pt signals very well with
a minimum $\chi^2$ of 70.5 for 85 effective degrees of freedom.
The derived one-dimensional marginalized credible interval for $S_8$ is
$0.76 \le S_8 \le 0.81$.
This result is consistent with the HSC-Y3 cosmic shear
studies \citep{Dalal_2023,Li_2023} but is $\sim 25$\% tighter than theirs.

We have not conducted a detailed comparison with the CMB results, as the
weak lensing results from the HSC final data will come soon.
In short, the results from HSC-Y3 cosmic shear studies
\citep{Dalal_2023,Li_2023} are in about $2\sigma$
tension with the {\it Planck} 2018 results \citep{2020A&A...641A...6P}.
Since our constraint is tighter than theirs, the tension should have
increased further. 

We have conducted a performance test of a photo-$3\times 2$-pt analysis
with mock catalogs that mimic the HSC-Y3 data in terms of the survey
area and redshift distribution of galaxies.
We have used mock galaxy catalogs constructed from full-sky
gravitational lensing simulations.
Mock galaxies are distributed according to a
linear bias relation with respect to the dark matter density field with
taking into account the lensing magnification effect.
The lensing shear, intrinsic alignment signal, and intrinsic shape noise
are assigned to each mock galaxy.
We have verified that unbiased estimates of the parameters of interest
can be obtained through a cosmological analysis with
mock HSC-Y3 photo-$3\times 2$-pt data (see Appendix \ref{apdx:mock} for
details).

Before closing the paper, we comment on differences between
photo-$3\times 2$-pt and the standard $3\times 2$-pt, and on a future
prospect.
The standard $3\times 2$-pt combines a spectroscopic galaxy catalog with
a photometric catalog.
The former is used as a lens sample, and the latter as a source
sample, from which three types of two-point statistics are derived:
galaxy clustering (spec$\times$spec), galaxy-galaxy lensing (spec$\times$photo)
and cosmic shear (photo$\times$photo).
Qualitatively speaking, it combines cosmological information from two
probes, galaxy-galaxy lensing supplemented by galaxy clustering and
cosmic shear, though they are correlated to some extent.
On the other hand, a photo-$3\times 2$-pt uses a single photometric galaxy
catalog for both lens and source samples.
It extracts information from two fields generated from
a single catalog, namely galaxy density fields, and weak lensing shear
fields, which are correlated to a greater extent.
This study demonstrates that a photo-$3\times 2$-pt enhances the
constraining power on particular nuisance parameters, thereby
improving $S_8$ constraints compared to a cosmic shear-only
analysis.
Therefore, the standard $3\times 2$-pt and
photo-$3\times 2$-pt analyses enhance the constraining power on $S_8$
through different paths, making them complementary to each other.

Recently, a joint analysis of them, referred to as $6\times 2$-pt,
is explored by \citet{Johnston_2025}.
They investigated the potential of using $6 \times 2$-pt to improve both
the precision and accuracy of the calibration of shear sample redshift
distributions, thereby reducing biases in the cosmological parameter
inference and increasing the statistical constraining power of the
surveys.
They performed simulated likelihood
  forecasts, and evaluated gains from the $6\times 2$-pt compared with  
  other 2-pt combinations.
  They found that while quantitative evaluations depend on specific
  analysis setups, qualitatively, the photo-$3\times 2$-pt 
  improves $S_8$ constraints compared to the cosmic shear-only
  case, though the $6\times 2$-pt yields further improvements.
In this paper, we have present the first application of the
photo-$3\times 2$-pt to real data, demonstrating the effectiveness of the
methodology, which provides partial empirical validation for $6 \times
2$-pt.
Given that the standard $3\times 2$-pt is now widely used
\citep[e.g.,][]{Miyatake_2023,Sugiyama_2023} and photo-$3\times 2$-pt
has also been put into practice, the $6 \times 2$-pt will become
a feasible and useful cosmological probe.

%
%
\begin{ack}
We would like to thank the anonymous referee for constructive comments
on the earlier manuscript which improved the paper.
We would like to thank M.~Shirasaki, T.~Kurita, M.~Oguri, M.~Takada, and
R.~Takahashi for useful discussions.
We would like to thank S.~Mineo for assistance with retrieving metadata
from HSC data archive system.
We would like to thank HSC data analysis software team for their effort
to develop data processing software suite, and HSC data archive team for
their effort to build and to maintain the HSC data archive system.
We are grateful to T.~Nishimichi, S.~Tanake, and
S.~Ishikawa for assistance with running
the $N$-body simulations with {\tt GINKAKU}.
We would like to thank Antony Lewis and 
Anthony Challinor for making the software {\tt CAMB} publicly available,
{\tt MultiNest} developers for {\tt MultiNest} publicly available, 
{\tt NaMaster} developers for {\tt Namaster} publicly available,
Robert Reischke for making the software {\tt OneCovariance} publicly available,
and 
{\tt HEALPix} team for {\tt HEALPix} software publicity available.

This work was supported in part by JSPS KAKENHI Grant
Number JP22K03655.

Data analysis were in part carried out on the analysis servers at Center for
Computational Astrophysics (CfCA), National Astronomical Observatory of
Japan (NAOJ). 
Numerical computations were in part carried out on Cray XC30 and XC50 at
CfCA, NAOJ, 
  
The Hyper Suprime-Cam (HSC) collaboration includes the astronomical
communities of Japan and Taiwan, and Princeton University.  The HSC
instrumentation and software were developed by NAOJ, the Kavli Institute
for the Physics and Mathematics of the Universe (Kavli IPMU), the
University of Tokyo, the 
High Energy Accelerator Research Organization (KEK), the Academia Sinica
Institute for Astronomy and Astrophysics in Taiwan (ASIAA), and
Princeton University.  Funding was contributed by the FIRST program from
the Japanese Cabinet Office, the Ministry of Education, Culture, Sports,
Science and Technology (MEXT), the Japan Society for the Promotion of
Science (JSPS), Japan Science and Technology Agency  (JST), the Toray
Science  Foundation, NAOJ, Kavli IPMU, KEK, ASIAA, and Princeton
University. 

This paper is based on data collected at the Subaru Telescope
and retrieved from the HSC data archive system, which is operated by
Subaru Telescope and Astronomy Data Center (ADC) at NAOJ. Data analysis
was in part carried out with the cooperation of CfCA at NAOJ.
We are honored and grateful for the
opportunity of observing the Universe from Maunakea, which has the
cultural, historical and natural significance in Hawaii. 

This paper makes use of software developed for Vera C. Rubin
Observatory. We thank the Rubin Observatory for making their code
available as free software at {\tt http://pipelines.lsst.io/}.

The Pan-STARRS1 Surveys (PS1) and the PS1 public science archive have
been made possible through contributions by the Institute for Astronomy,
the University of Hawaii, the Pan-STARRS Project Office, the Max Planck
Society and its participating institutes, the Max Planck Institute for
Astronomy, Heidelberg, and the Max Planck Institute for Extraterrestrial
Physics, Garching, The Johns Hopkins University, Durham University, the
University of Edinburgh, the Queen’s University Belfast, the
Harvard-Smithsonian Center for Astrophysics, the Las Cumbres Observatory
Global Telescope Network Incorporated, the National Central University
of Taiwan, the Space Telescope Science Institute, the National
Aeronautics and Space Administration under grant No. NNX08AR22G issued
through the Planetary Science Division of the NASA Science Mission
Directorate, the National Science Foundation grant No. AST-1238877, the
University of Maryland, Eotvos Lorand University (ELTE), the Los Alamos
National Laboratory, and the Gordon and Betty Moore Foundation. 
\end{ack}

%
%
\begin{appendix}
%
%
\section{Performance test with mock catalogs}\label{apdx:mock}

We conduct a performance test of a photo-$3\times 2$-pt analysis with mock
catalogs.
The primary aim of this test is to verify whether a parameter inference
with photo-$3\times 2$-pt data under realistic survey settings returns
unbiased estimates of the parameters of our interest.

Below, we first describe numerical simulations to create mock catalogs
for photo-$3\times 2$-pt measurements, then we present results from cosmological
analyses of mock data vectors.

%
%
\subsection{Numerical simulations}\label{apdx:simulations}

%
%
\begin{table*}
\caption{$N$-body simulation parameters. \label{table:nbody} } 
\begin{tabular}{llccl}
\hline
Box size & Redshift coverage & Particle mass & Minimum halo mass & Dumped redshift \\
$[h^{-1}$Mpc$]$ & {} & $[h^{-1}M_\odot]$ & $[h^{-1}M_\odot]$ $(N_{\rm particle})$ & {} \\
\hline
545 & $0 \le z< 0.191$ & $1.7\times10^{9}$ & $1.0\times10^{12}$ (605) &
0.010, 0.031, 0.052, 0.075, 0.099, 0.124, 0.149, 0.176 \\
1150 & $0.191 \le z< 0.429$ & $1.6\times10^{10}$ & $9.9\times10^{11}$ (64) &
0.205, 0.235, 0.266, 0.299, 0.333, 0.370, 0.408 \\
1790 & $0.429 \le z< 0.724$ & $5.8\times10^{10}$ & $2.9\times10^{12}$ (50) &
0.449, 0.923, 0.538, 0.587, 0.639, 0.695 \\
2430 & $0.724 \le z< 1.083$ & $1.5\times10^{11}$ & $7.3\times10^{12}$ (50) &
0.754, 0.818, 0.887, 0.961, 1.041 \\
3020 & $1.083 \le z< 1.500$ & $2.8\times10^{11}$ & $1.1\times10^{13}$ (40) & 1.128, 1.222, 1.326,
1.439 \\
3700 & $1.500 \le z< 2.125$ & $5.2\times10^{11}$ & $2.1\times10^{13}$ (40) & 1.564, 1.703, 1.857,
2.030 \\
4490 & $2.125 \le z< 3.167$ & $9.2\times10^{11}$ & $2.8\times10^{13}$ (30) & 2.226, 2.448, 2.704,
3.000 \\
5440 & $3.167 \le z< 5.250$ & $1.6\times10^{12}$ & $4.9\times10^{13}$ (30) & 3.345, 3.762, 4.263,
4.882 \\
\hline
\end{tabular}
\end{table*}

Our methodology of the gravitational lensing ray-tracing simulation
largely follows that of \citet{2017ApJ...850...24T} \citep[see
  also][]{Shirasaki_2015} which we refer the reader to, but there are
some differences that we describe in the following subsections.

The overview of numerical simulations is as follows:
Mass distributions in the Universe are constructed by
multi-layer shells of dark matter.
The boundaries of the shells are set by the radial distances corresponding to
a constant interval of the cosmic scale factor $\Delta a=0.02$.
Dark matter distributions are generated by series of $N$-body
simulations with different box sizes, and dark matter particles
within each shell are dumped at a single time, see appendix~\ref{apdx:nbody}).
Dark matter particles in each shell are projected onto lens planes, and
the generated surface mass density maps are used to generate lensing
convergence maps and then shear maps (see appendix~\ref{apdx:ray-tracing}). 
The surface mass density maps are also used to generate intrinsic
alignment fields for which we adopt the linear alignment model (see
appendix~\ref{apdx:IA}).

For the cosmological numerical simulations, we adopt the standard
$\Lambda$ cold dark matter 
($\Lambda$CDM) cosmology that is consistent with the Planck CMB data
\citep{2016A&A...594A..13P}: The cosmological parameters are, the matter 
density parameter $\Omega_{\rm m}=0.316$, the baryon density
$\Omega_{\rm b}=0.049$, the cosmological constant $\Omega_\Lambda=0.684$,
the Hubble parameter $h=0.672$, the amplitude of density fluctuations
$\sigma_8=0.83$, and the spectral index $n_{\rm s}=0.965$.
For the sum of neutrino mass, we take $\Sigma m_\nu = 0.06$eV.

\subsubsection{$N$-body simulation}\label{apdx:nbody}


We conducted a series of cosmological $N$-body simulations on the
periodic cubic box following the gravitational evolution of dark matter
particles without baryonic processes.
We adopted the cosmological $N$-body simulation code {\tt GINKAKU}
\citep{https://doi.org/10.48550/arxiv.2605.28581}. 
This code employs the Tree Particle-Mesh (TreePM) method to compute the
gravitational force with the short-range tree force being implemented
based on the Framework for Developing Particle Simulators 
\citep[FDPS;][]{Iwasawa_2016,Namekata_2018}, a public library for general
particle simulations.

We used simulation boxes with 8 different side lengths (from 545 to
5440$h^{-1}$Mpc) to cover upto a redshift of $z=5.25$ over the full-sky.
The number of particles for each box was $2048^3$.
The particle positions and velocities were dumped at redshifts
corresponding to the cosmic expansion factors of, $a_i=0.99 -0.02 \times
(i-1)$ ($i=1,2,3,,,42$).
See Table \ref{table:nbody} for settings of $N$-body simulations.
Since the simulation settings are similar to those of
\citet{2017ApJ...850...24T}, the properties of the simulation products
(lensing maps and dark matter halo samples) are also similar.
We have run 16 sets of $N$-body simulations of 8 different box sizes.

\subsubsection{Gravitational lensing ray-tracing simulation}\label{apdx:ray-tracing}
Dark matter particles in each shell are projected onto a lens plane
spherical surface.
We used {\tt HEALPix} pixelization \citep{Gorski_2005},
with {\tt HEALPix} parameter of $N_{\rm side}=4096$ or the pixel size of
$\theta_{\rm pix}=51.5\arcsec$, and we generated the projected surface
mass density fluctuation maps by assigning each particle to the nearest pixel,
denoted by $\delta \Sigma^i$ (where $i$ is the index of a lens plane).
Then we have the convergence field for $i$th shell
\citep[see][for details]{Shirasaki_2015},
\begin{equation}
\label{eq:K}
K^i(\bm{\theta})={{4\pi G} \over c^2}
{{\delta \Sigma^i(\bm{\theta})} \over {a_{L,i} D_j}},
\end{equation}
where $a_{L,i}$ is the scale factor of the $i$th lens plane, which we
take $a_{L,i}=0.99-0.02 \times (i-1)$, and $D_i$ is the radial comoving
distance to the lens plane.

In computing the lensing convergence field, we adopt Born approximation
\citep{Ferlito_2024},
instead of the multiple-lens plane algorithm adopted in
\citet{2017ApJ...850...24T}.
The lensing convergence field of the $j$th source plane is given by,
\begin{equation}
\label{eq:kappa}
\kappa^j(\bm{\theta})=\sum_{i=1}^j {{D_{ji}} \over {D_j}}K^i(\bm{\theta})
\end{equation}
where $D_{ji}$ is the radial comoving
distance between the $i$th lens plane and $j$th source plane.
We set 42 source planes placed at the far boundaries of each shell with
the corresponding scale factor being $a_{S,j}=1-0.02 \times j$.
The lensing shear field, $\gamma(\bm{\theta})$, is obtained from the
convergence field via the relation in the spherical harmonics space
\citep{Hu_2000},
\begin{equation}
\label{eq:shear}
\gamma_{\ell m}=-\sqrt{{(\ell -1)(\ell+2)}\over{\ell(\ell+1)}}
\kappa_{\ell m}.
\end{equation}

The lensing convergence and shear fields generated from the multi-layer
shells is known to be affected by the finite-shell-thickness effect,
which arises due to the lack of radial-mode fluctuations on scales
larger than a shell-thickness \citep[see Appendix B
  of][]{2017ApJ...850...24T}. 
We correct this in the following manner:
We compute theoretical power spectra of the surface mass density field of a
shell with and without the finite-shell-thinness effect by using equations
(25) and (26) of \citet{2017ApJ...850...24T}, and denote them as
$C_\ell^{\textrm{shell-fst}}$ and $C_\ell^{\rm shell}$, respectively.
We introduce a correction term,
$C_{\rm fst}(\ell)=C_\ell^{\rm shell}/C_\ell^{\textrm{shell-fst}}$, 
and apply it to the spherical harmonics components of $K^i(\bm{\theta})$
as
\begin{equation}
\label{eq:fst}
K_{\ell m}^i \rightarrow \sqrt{C_{\rm fst}(\ell)} K_{\ell m}^i.
\end{equation}
We compared the resulting convergence power spectra from simulation maps with the
theoretical prediction, and confirmed that those agree within 3-percents
over the range $50 \lesssim \ell \lesssim 1000$ for $z_s>0.3$.
It should be noticed that this correction is reasonable for two-point
statistics but not for other statistics.

In order to increase the number of realizations, we choose 16
observer's positions for each box by using the maximum distance sliced Latin
hypercube designs algorithm \citep{Ba_2015}, which enables 
random and homogeneous sampling in a high-dimensional parameter space. 
Since we have generated 16 $N$-body realizations,
we have $16\times 16 =256$ sets of shells for each $N$-body box size.
Then we randomly combined them to have 256 sets of full-sky lensing
simulation data.
We also generated dark matter halo samples in the same manner as
\citet{2017ApJ...850...24T}.

\subsubsection{Linear intrinsic alignment from tidal field}\label{apdx:IA}

We adopt the non-linear alignment (NLA) model
\citep{Hirata_2004,2007NJPh....9..444B}.
This model is considered one of the origins of intrinsic alignment
phenomenon and is widely used in weak lensing studies
\citep[for reviews, see e.g.][]{2015PhR...558....1T,Chisari_2025},
making it suitable for performance testing of the photo-$3\times 2$-pt.
In the NLA model, intrinsic alignments of galaxies are caused by a tidal
field at the galaxy positions.
We use the two-dimensional tidal field of the projected surface density
field, $\delta {\Sigma}^i$, instead of the three-dimensional tidal field.
This is a practical method for simplifying computations, but it yields
IA angular correlation signals consistent with those predicted by 
the NLA model \citep{Harnois_D_raps_2021}.

The procedure to compute IA maps from the projected surface density
maps is as follows.
First, we compute a spin-2 tidal fields, $s^{\rm IA}$, of
$\delta {\Sigma}^i$ via the spherical harmonics space using the same
relation as equation (\ref{eq:shear}). 
Then the NLA intrinsic alignment ellipticity map for $i$th shell is
given by,
\begin{equation}
\label{eq:eIA}
\epsilon_{1/2}^i(\bm{\theta}) = -{{A_1^i C_1 \Omega_m, \rho_{\rm cr,0}} \over {D_+(z_i)}}
{{s_{1/2}^{\rm IA}}(\bm{\theta}) \over {\Delta\chi_i}},
\end{equation}
where $A_1^i$ is the model parameter controlling the relation
between a tidal field and IA ellipticity,
$C_1=5\times 10^{-14}M_\odot^{1} h^{-2}{\rm Mpc}^3$
is a conventional constant, $\rho_{\rm cr,0}$ is the critical density
today, $D_+(z)$ is the linear growth factor normalized to unity today,
and $\Delta\chi_i$ is the comoving thickness of the $i$th shell.
These pixel values of the IA maps are used to assign IA signals to mock
galaxies. Since mock galaxies follow a linear bias
relation with respect to the dark matter density field and are thus
spatially clustered, the resulting IA-related power spectra
inherently include the density-weighting effect. (see section
\ref{ssec:Angularpowerspecrtra}).

\subsection{Mock galaxy catalog and power spectrum measurement}\label{apdx:mockgalaxy}

Using the simulation data set described above, we
created mock galaxy catalogs fol
lowing the procedure below:
The basic design of the mock catalog is based on the HSC-Y3 weak lensing
galaxy catalog \citep{Li_2022}.
We defined a survey region covering $\sim418$ sq degrees on
the equator, with \(\Delta\text{R.A.} = 21^\circ\) and
\(\Delta\text{Dec} = 20^\circ\).
This coverage is roughly equivalent to the HSC-Y3 footprint.

We generate tomographic galaxy samples with four redshift bins. 
The numbers and redshift distributions of galaxies are taken from those
of HSC-Y3 cosmic shear analyses \citep{Dalal_2023,Li_2023}, but redshift
distributions are re-binned from the original redshift binning of
$\Delta z_{\rm bin}=0.025$ to the redshift intervals of shells, corresponding to
$\Delta a_i = 0.02$. 
In each shell, sky positions of mock galaxies are randomly assigned
according to a linear relation with the density fluctuations, which
arise from the projected matter distribution and
gravitational lensing magnification ($\delta_\mu^i$, computed from the
lensing convergence and shear field), namely $\delta_g^i = \delta
\Sigma^i + \delta_\mu^i$.
Thus the generated galaxy samples have a galaxy clustering bias of
$b_g=1$.
We set the magnification parameters to $\alpha_\mu^i=(0.3, 0.3, 0.7, 1.2)$
for 4 tomographic samples (from the lowest to the highest bin), which
are consistent with those measured from HSC-Y3 galaxy catalog (see
section~\ref{sssec:magnifparams}). 

For each mock galaxy in a shell, we assign the lensing shear and
intrinsic alignment ellipticities taken from the nearest pixel of
the corresponding simulation maps.
We set the NLA model parameter $A_1^i=0.5$ for all the 4 tomographic
samples.
Finally, for each mock galaxy, we assign the galaxy intrinsic
ellipticity and weight taken randomly from the real HSC-Y3 galaxy
catalog. 
We have created 64 sets of mock galaxy catalogs.

\subsection{Power spectrum measurement}\label{apdx:mock-measurement}

Measurements of power spectra of shear-shear, galaxy-shear, and
galaxy-galaxy were done in the same manner as in the real data analyses
(section \ref{sec:measurements}).
Briefly, we used the pseudo-$C_\ell$ method implemented in
{\tt NaMaster} \citep{Alonso_2019}.
The shear and galaxy density maps are in {\tt HEALPix} pixelization
format \citep{Gorski_2005} with {\tt HEALPix} parameter of
$N_{\rm side}=4096$.
Weight maps are the sum of weight \citep{Nicola_2021} for shear maps and
binary mask for galaxy density maps.
No area cut by observational conditions was made for galaxy density map
as none of variations of observational conditions were considered in
creating mock data.
Note that since we define the mock survey region by all {\tt HEALPix}
pixels whose
pixel center is within an area of $\Delta$R.A.$=21\degree$
and $\Delta$Dec$=20\degree$, all the pixels are fully covered by
the survey area.

\subsection{Parameter inference}\label{apdx:mock-infarence}

We employ the standard Bayesian likelihood analysis for
the parameter inference of measured data vector, which has the same
content as those in the real HSC-Y3 photo-$3\times 2$-pt analysis (see
section~\ref{ssec:datavec}).
The log-likelihood is computed employing the Gaussian covariance matrix, 
which is computed with {\tt NaMaster} \citep{Alonso_2019} using the
narrow kernel approximation (NKA) estimator developed in
\citet{Garc_a_Garc_a_2019}.
Note that the connected non-Gaussian and super-sample contributions to
the covariance matrix, which are neglected in this mock analysis, are
subdominant over our $\ell$-ranges but are not negligible. 
Therefore the error bars presented below might be
slightly underestimated. 

Parameter inference was done in the same manner as that of the 
HSC-Y3 data analysis, using a publicly available software {\tt MultiNest}
\citep{2008MNRAS.384..449F,2009MNRAS.398.1601F,2019OJAp....2E..10F}. 
We adopt the same set of model parameters as in the real HSC-Y3 data
analysis (see section~\ref{ssec:modelparams}), except that the baryonic
feedback parameter is fixed to be 
$A_{\rm bary}=3.13$ as no baryonic effect is included in the simulations.

\subsection{Results of mock analyses}\label{apdx:mock-results}

\begin{figure}
\begin{center}
\includegraphics[width=82mm]{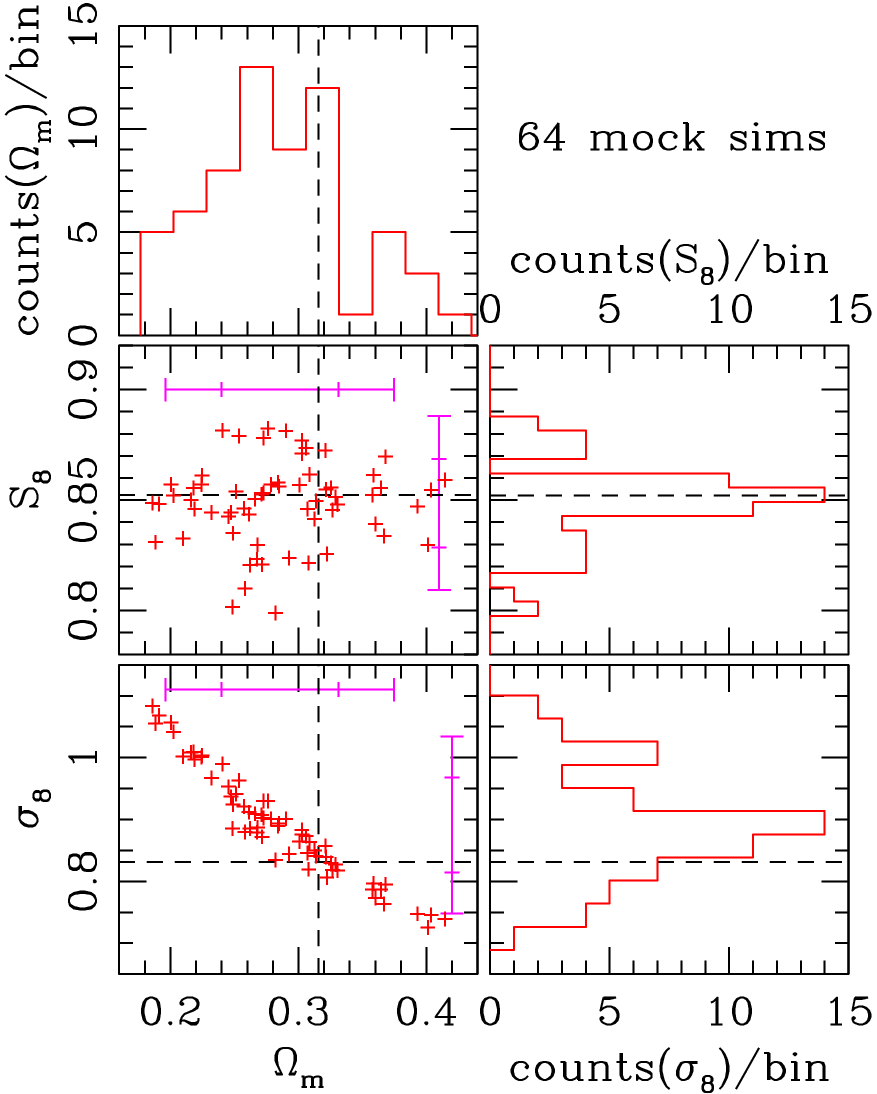}
\end{center}
\caption{Middle left and bottom left panels: Scatter plots showing inferred
  values (means of marginalized one-dimensional posterior distributions)
  in two parameter spaces of $\Omega_m$-$S_8$ and $\Omega_m$-$\sigma_8$.
  Error bars show averaged 68- and 95-percents credible intervals
  centered at averaged means over 64 realizations.
  Top and right panels: Frequency distributions of inferred
  mean values. The dashed lines show the input parameter values used in
  creating mock catalogs.  
  \label{apdx:fig:mock_om_sig8_s8}}
\end{figure}

\begin{figure}
\begin{center}
\includegraphics[width=82mm]{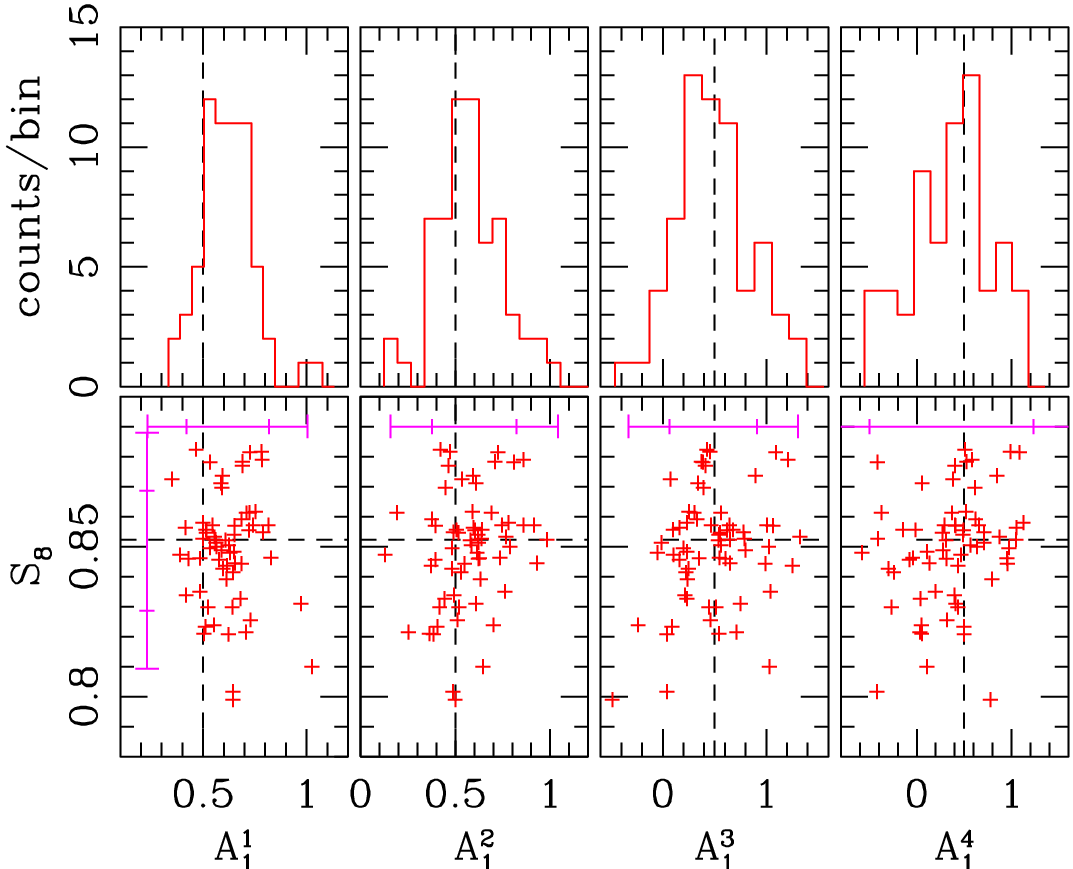}
\end{center}
\caption{Bottom panels: Scatter plots showing inferred
  values in $A_1^i$-$S_8$ plane. Error bars
  show averaged 68- and 95-percents credible intervals centered at
  averaged means over 64 realizations.
  Top panels: Frequency distributions of inferred values.
  The dashed lines show the input parameter values used in
  creating mock catalogs.  
  \label{apdx:fig:mock_aone}}
\end{figure}

Here we present results of our mock analyses using two types of
plots showing distributions of inferred values of model parameters,
presented in Figures~\ref{apdx:fig:mock_om_sig8_s8}-\ref{apdx:fig:mock_alpmag}.
One is histograms showing frequency distributions of inferred values.
The other is scatter plots showing distributions of
inferred values in two parameter spaces.
For the inferred value, we take the mean of a marginalized
one-dimensional posterior distribution.
In the scatter plots, we show representative error bars, which are 
averaged 68- and 95-percents credible intervals centered at averaged
means over 64 realizations.

%
%

\subsubsection*{Cosmological parameters}
Figure~\ref{apdx:fig:mock_om_sig8_s8} shows distributions of inferred
values of key cosmological parameters ($\Omega_m$, $\sigma_8$ and
$S_8$).
It is found from this Figure that unbiased estimates for those
cosmological parameters can be obtained from the photo-$3\times 2$-pt
analysis. 

%
%

\subsubsection*{IA parameters}
Results for IA amplitude parameter $A_1^i$ are shown in Figure~\ref{apdx:fig:mock_aone}, in
which it is found that unbiased estimates for those parameters are obtained.
Note that due to the IA parameter renormalization \citep[a procedure in
  the perturbative expansion where
contributions from small-scale physics to large-scale correlations are
absorbed into effective IA parameters, see][]{Blazek_2019}, the inferred IA
parameters ($A_1^i$) do not necessarily equal the input values. 
Other IA parameter, $b_T^i$, are poorly constrained because IA terms
involving this parameter is subdominant. In other words, $b_T^i$ has no
strong impact on inference of other parameters.

%
%
\begin{figure}
\begin{center}
\includegraphics[width=82mm]{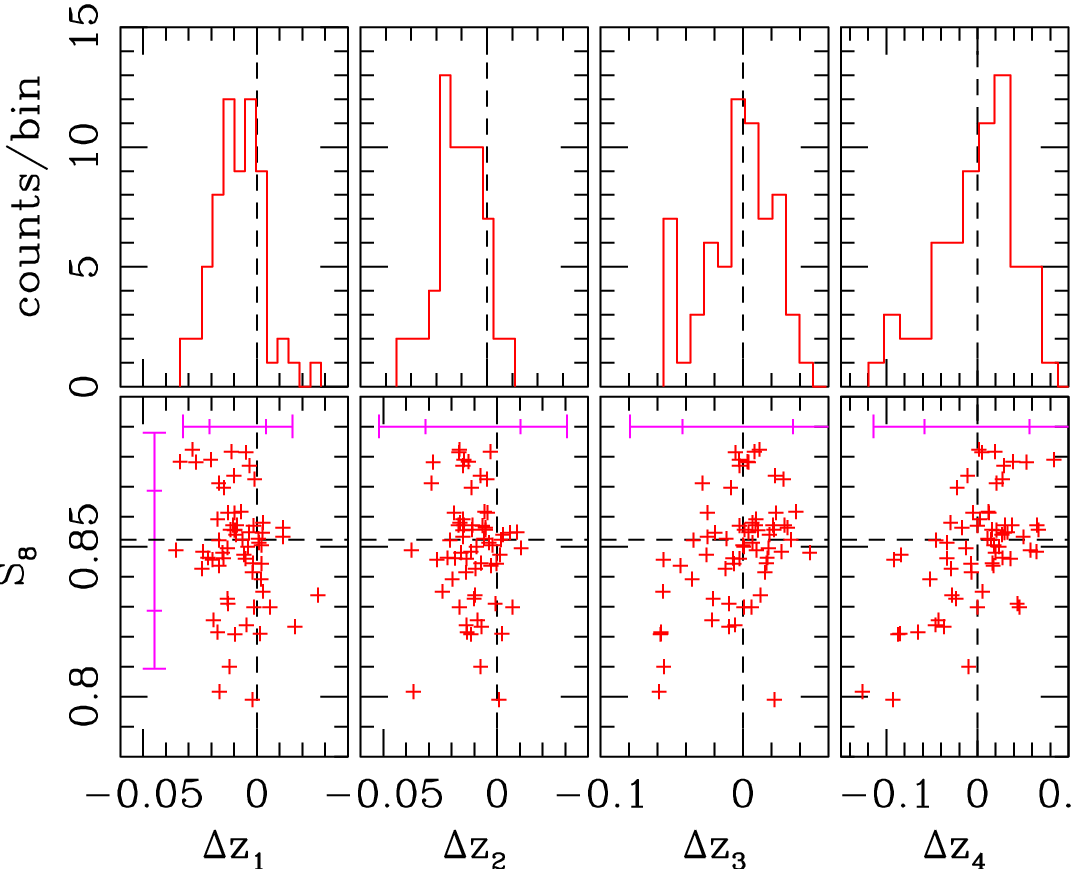}
\end{center}
\caption{Same as Figure~\ref{apdx:fig:mock_aone} but for the shift
  parameters of the galaxy redshift distributions, $\Delta z_i$.  
  \label{apdx:fig:mock_delz}}
\end{figure}

\subsubsection*{$p(z)$ shift parameter}
Results for the shift parameters ($\Delta z_i$) of galaxy redshift
distributions ($p(z)$)  are shown in Figure \ref{apdx:fig:mock_delz}.
Note that for $\Delta z_1$ and $\Delta z_2$, the Gaussian prior with
$\sigma=0.03$ is adopted, and thus the results for those two parameters
are prior dominated.
On the other hand a flat prior with $-0.5<\Delta z_i<0.5$ is adopted for
$\Delta z_3$ and $\Delta z_4$, and thus the results demonstrate that
unbiased estimates for those parameters can be obtained 
from the photo-$3\times 2$-pt analysis. 

%
%
\begin{figure}
\begin{center}
\includegraphics[width=82mm]{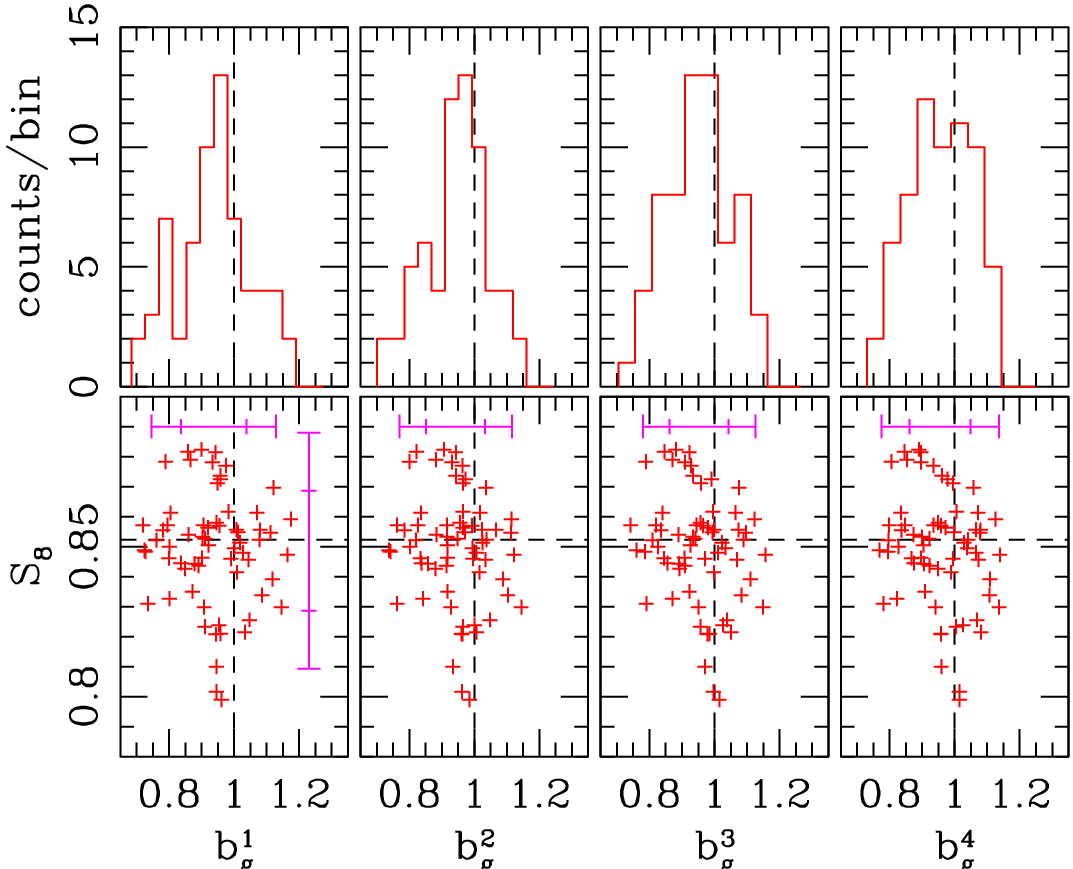}
\end{center}
\caption{Same as Figure~\ref{apdx:fig:mock_aone} but for the galaxy
  clustering bias parameter $b_g^i$.  
  \label{apdx:fig:mock_bg}}
\end{figure}

\subsubsection*{Galaxy clustering bias parameter}
Results for the galaxy clustering bias parameter ($b_g^i$) are shown in
Figure~\ref{apdx:fig:mock_bg}.
It is found from the figure that unbiased estimates for the bias
parameters can be obtained.

%
%
\begin{figure}
\begin{center}
\includegraphics[width=82mm]{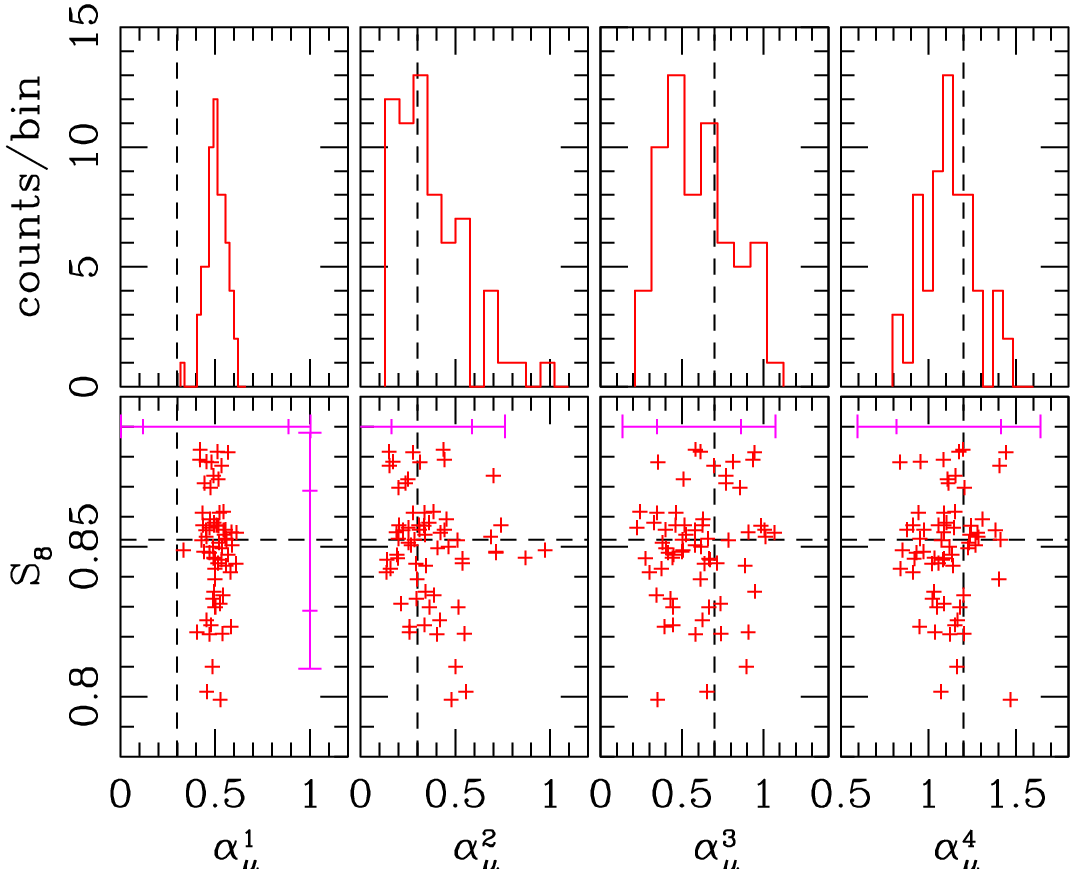}
\end{center}
\caption{Same as Figure~\ref{apdx:fig:mock_aone} but for the
  lensing magnification parameter $\alpha_\mu^i$.  
  \label{apdx:fig:mock_alpmag}}
\end{figure}

\subsubsection*{Lensing magnification parameter}
Results for the lensing magnification parameter ($\alpha_\mu^i$) are shown in
Figure~\ref{apdx:fig:mock_alpmag}. 
Regarding the result for $\alpha_\mu^1$, it is found that the inferred
values are systematically shifted towards larger values with very
large error bar.
This is because the magnification effect is negligibly small for this
lowest redshift galaxy sample, so no useful constraint is placed on
$\alpha_\mu^1$. 
For other cases, it is found that unbiased estimates for those
parameters are obtained.

%
%
\section{Area cuts of galaxy density maps to
  mitigate effects of systematics}\label{apdx:systematics}

%
\begin{table}
\caption{Summary of area cuts of galaxy density maps by
  systematics. \label{table:systematics}} 
\begin{tabular}{lcc}
\hline
Systematics & Threshold & Removed area \\
\hline
Survey depth & $\ge 25.5$ & 3.0\% \\
Dust extinction & $\le 0.15$ & 3.8\% \\
Star number density & No cut & {} \\
Sky level & $ \le 4000$ & 2.2\% \\
Sky variance & $\ge 0.003$ & 0.4\% \\
Mean airmass & $\le 1.5$ & 1.1\% \\
Seeing & $\ge 0.5$ and $\le 0.75$  & 10.2\% \\
Total exposure time & $\ge 800$ & 0.5\% \\
Number of input frames & $\ge 4$ & 1.6\% \\
Pixel-coverage rate & $\ge 0.8$ & 12.9\% \\
\hline
Total & {} & 29.5\% \\
\hline
\end{tabular}
\end{table}

\begin{figure}
\begin{center}
\includegraphics[width=82mm]{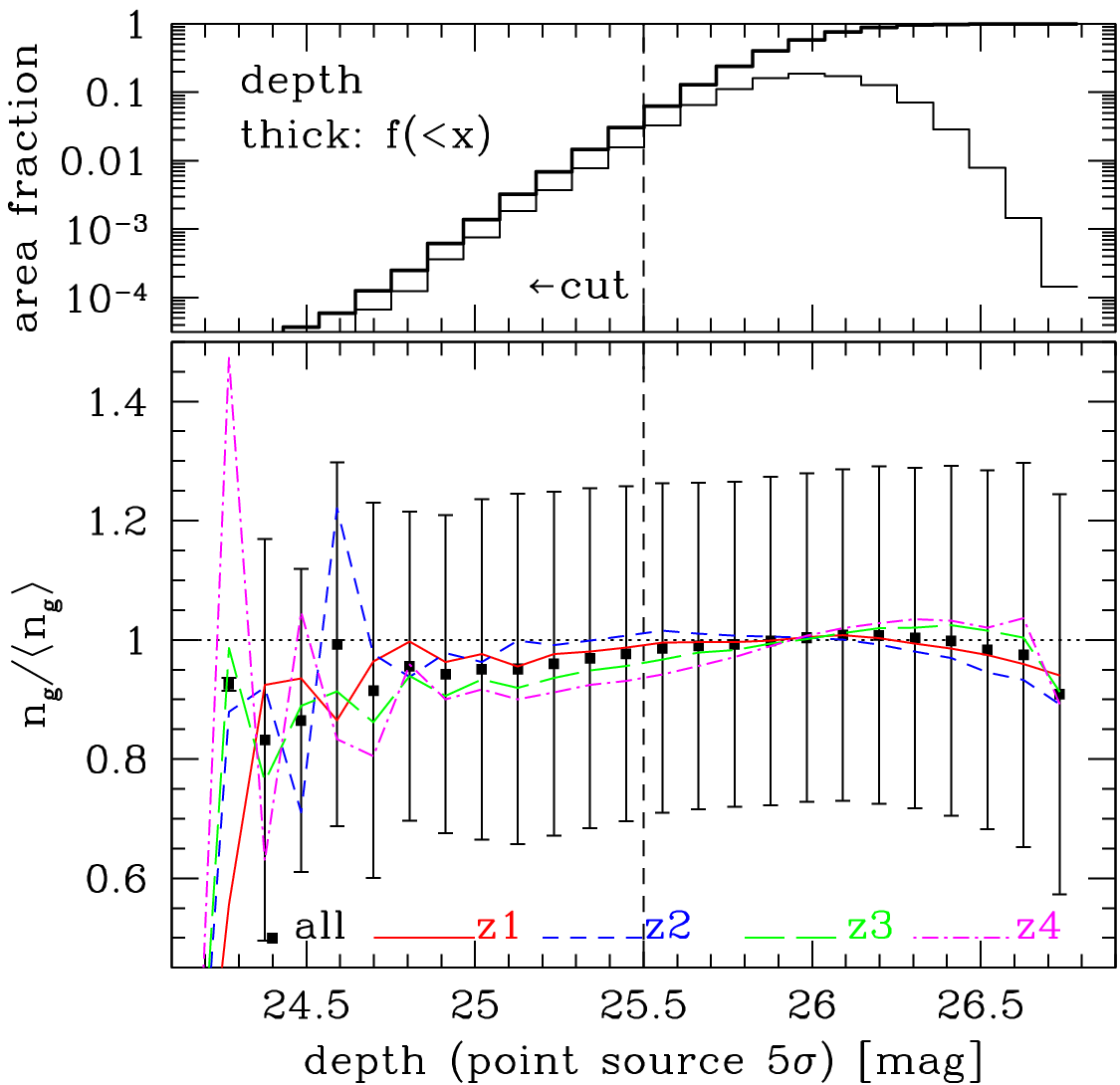}
\end{center}
\caption{Top panel: The area fraction as a function of $5\sigma$ point
  source survey depth. The thin and thick lines are for the differential
  and accumulated fractions, respectively.
  Bottom panel: Filled squares with error bars show the mean number
  density of galaxies from all four
  tomographic samples as a function of binned survey depth, normalized
  by the mean galaxy number density over all the bins.
  The galaxy density maps and systematic map with {\tt HEALPix}
  resolution parameter $N_{\rm side}=2048$ are used.
  Error bars show the root-mean-square of galaxy
  densities over pixels within each bins.
  The size of the error bars is predominantly determined by galaxy
  clustering, and therefore exhibits little dependence on the $x$-axis
  value. 
  The lines shows the normalized mean galaxy number densities for each
  tomographic samples with the solid, dashed, long-dashed, and
  dot-dashed lines being for $z_1$ (the lowest $z$-bin), $z_2$,  $z_3$,
  and $z_4$ (the highest $z$-bin) samples, respectively.  
  \label{apdx:sys:fig:sdepth}}
\end{figure}

\begin{figure}
\begin{center}
\includegraphics[width=82mm]{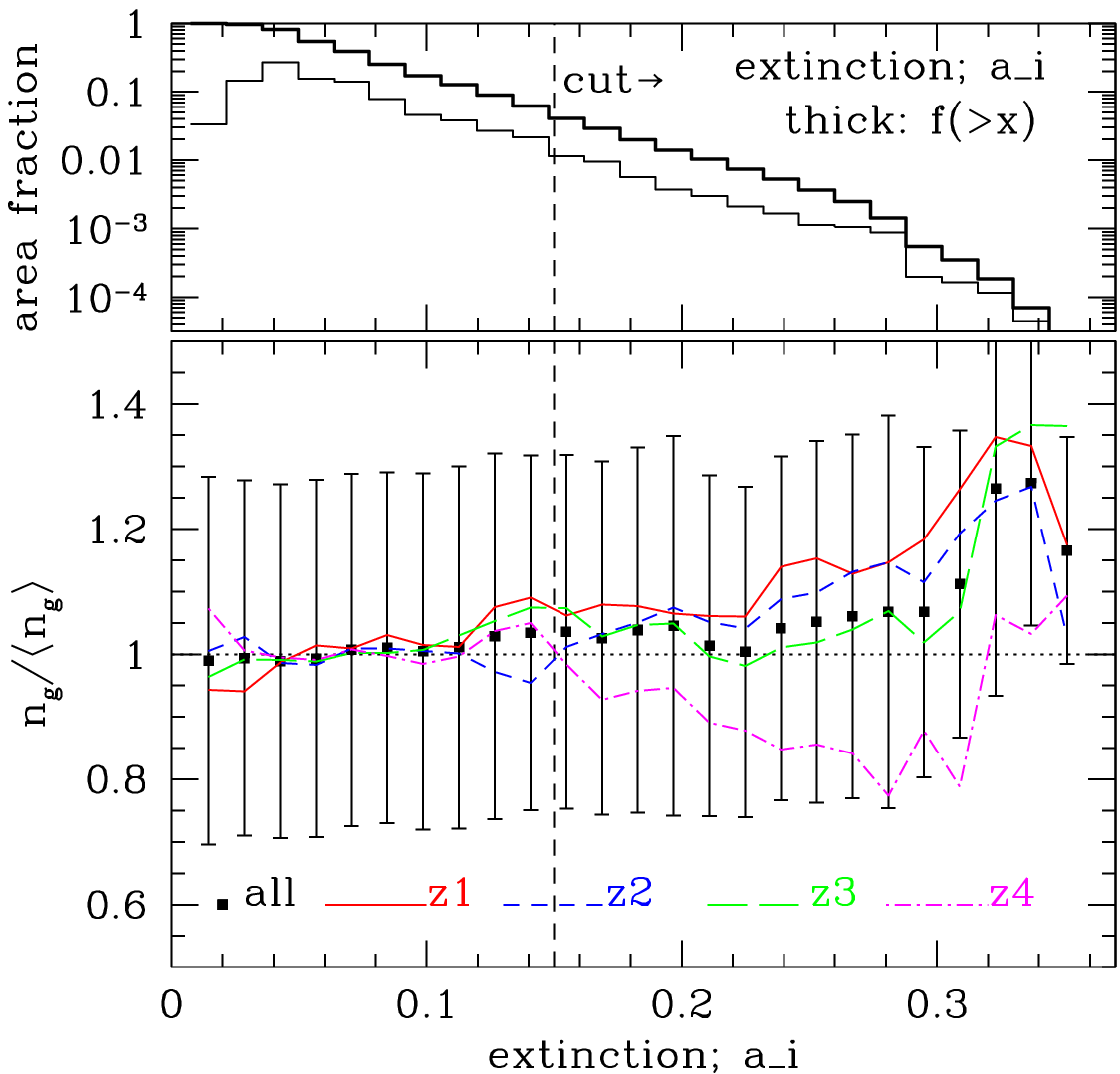}
\end{center}
\caption{Same as Figure~\ref{apdx:sys:fig:sdepth} but for the dust
  extinction.  
  \label{apdx:sys:fig:extinc}}        
\end{figure}

\begin{figure}
\begin{center}
\includegraphics[width=82mm]{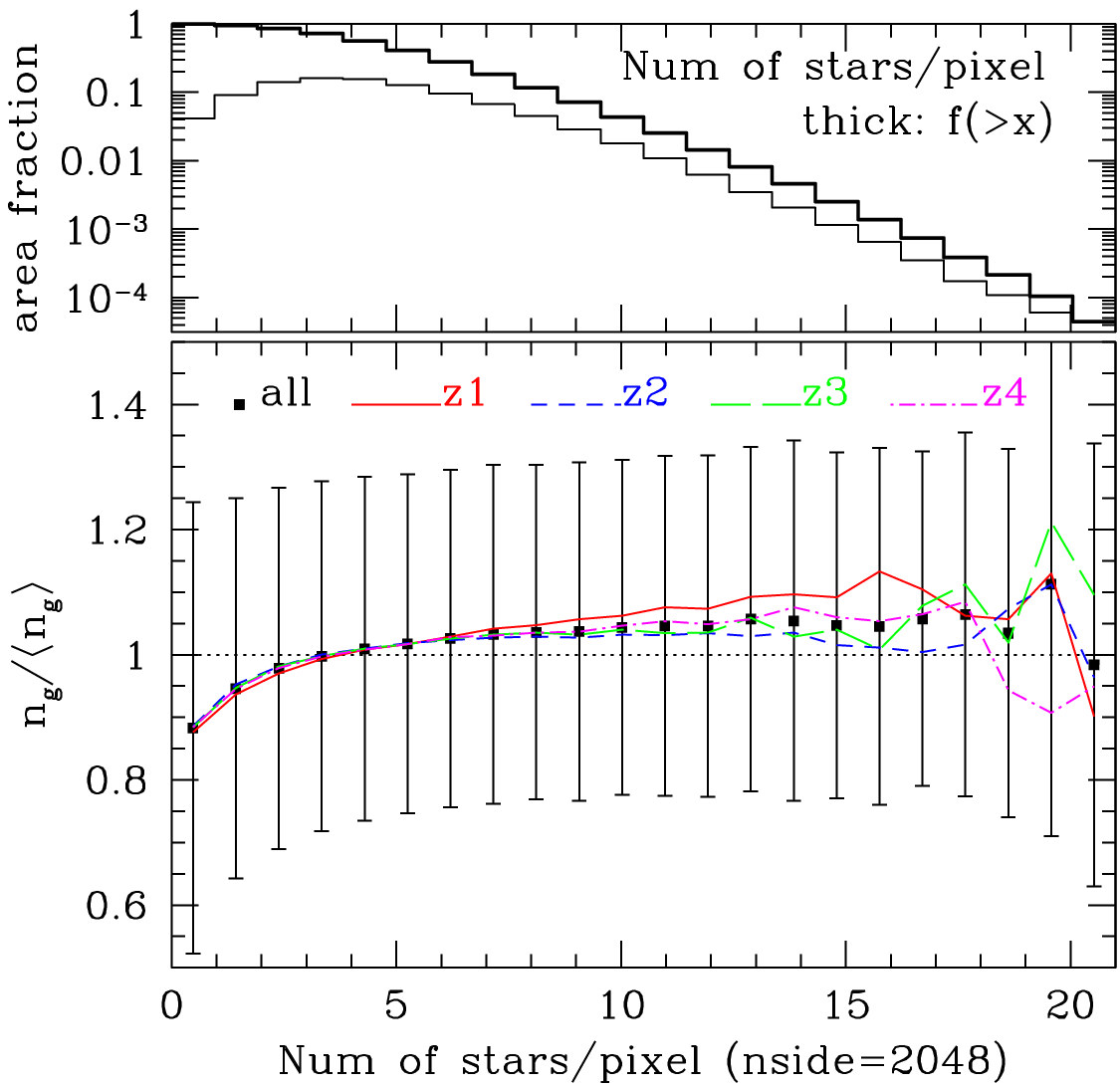}
\end{center}
\caption{Same as Figure~\ref{apdx:sys:fig:sdepth} but for the number of
  stars in each pixel with the {\tt HEALPix} parameter of $N_{\rm
    side}=2048$. No cut was imposed for this case.  
  \label{apdx:sys:fig:nstar}}
\end{figure}

\begin{figure}
\begin{center}
\includegraphics[width=82mm]{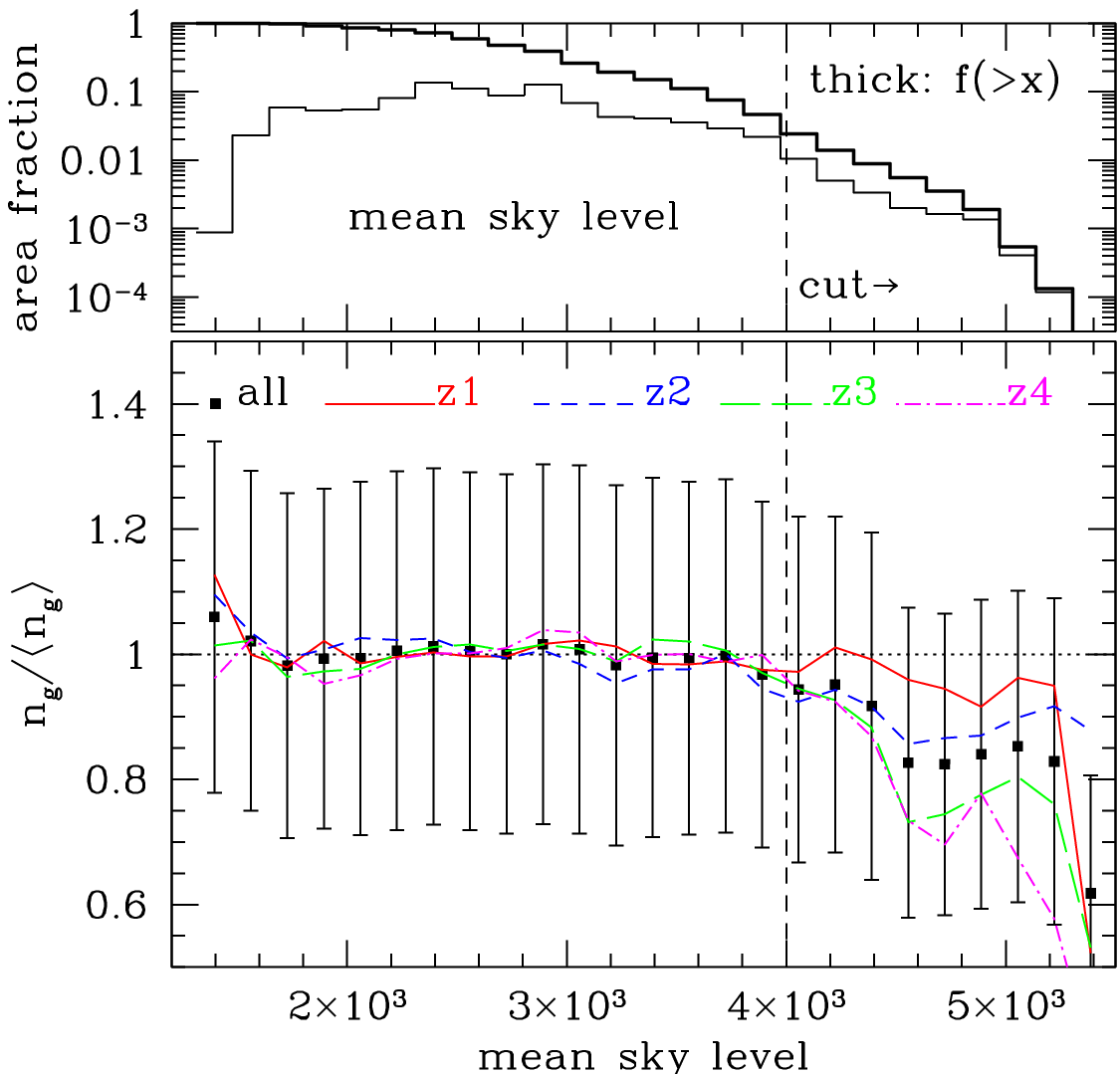}
\end{center}
\caption{Same as Figure~\ref{apdx:sys:fig:sdepth} but for the mean sky
  level.  
  \label{apdx:sys:fig:skylev}}
\end{figure}

\begin{figure}
\begin{center}
\includegraphics[width=82mm]{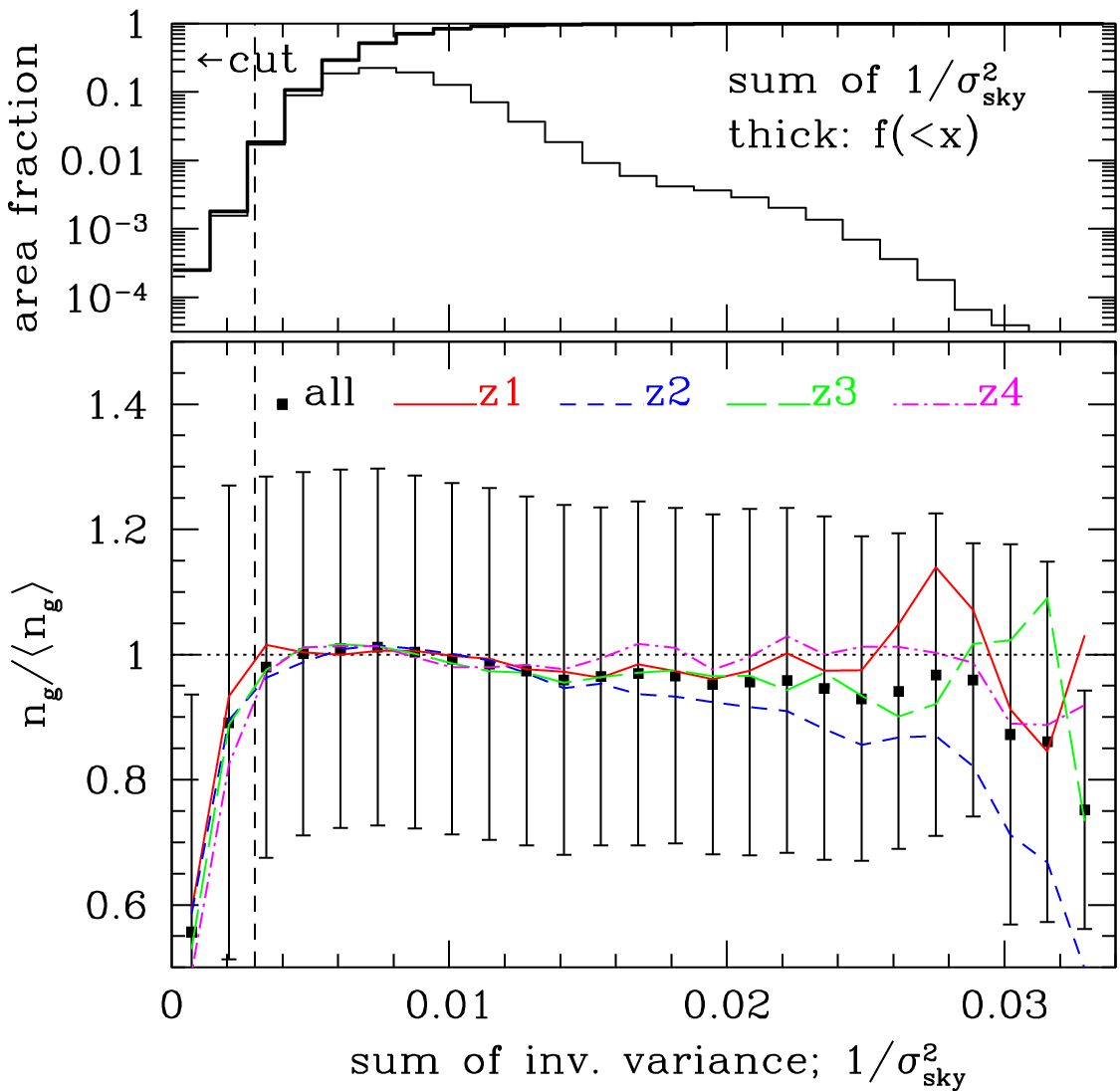}
\end{center}
\caption{Same as Figure~\ref{apdx:sys:fig:sdepth} but for the inverse of 
  sky variance.  
  \label{apdx:sys:fig:varsky}}
\end{figure}

\begin{figure}
\begin{center}
\includegraphics[width=82mm]{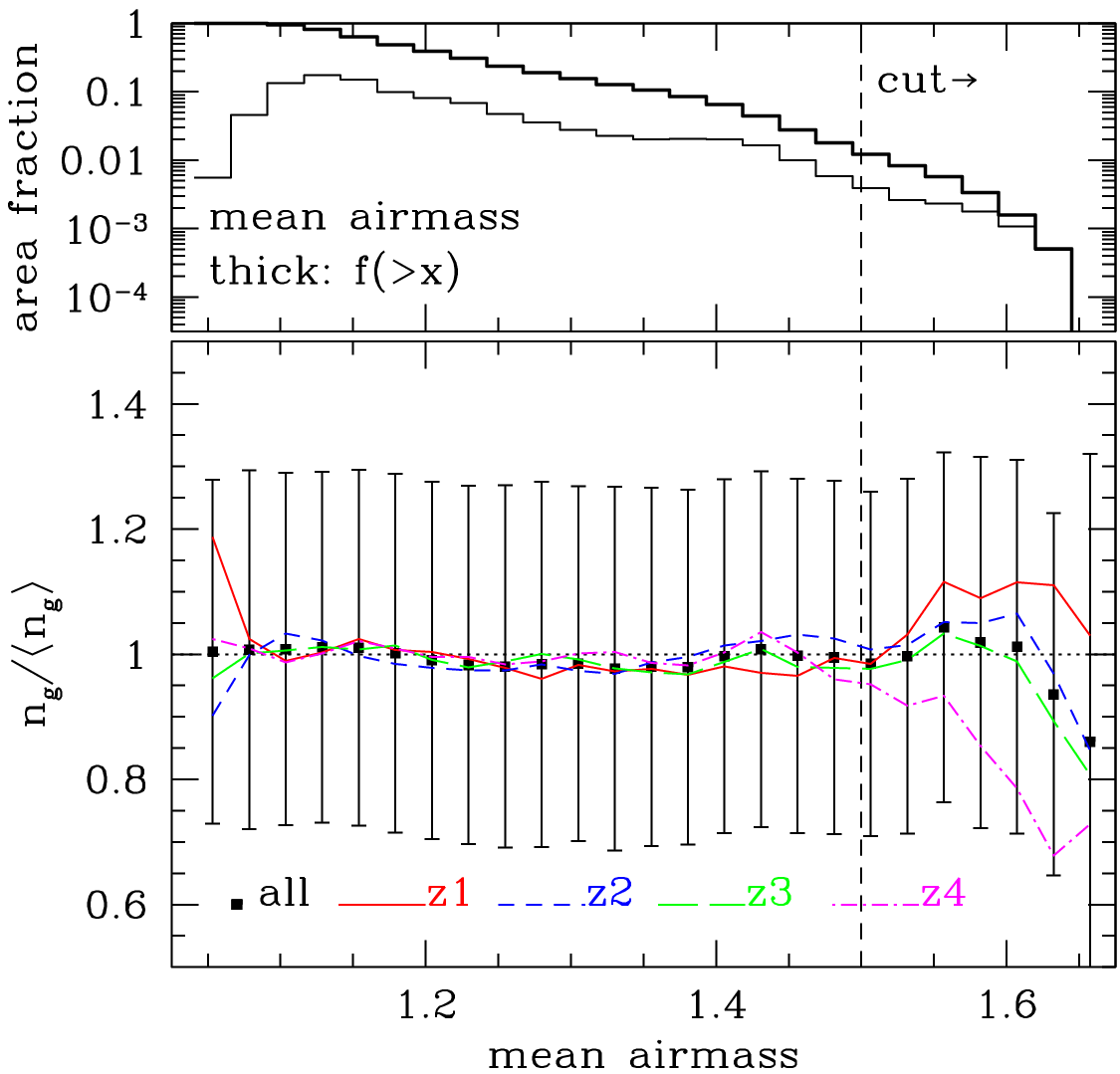}
\end{center}
\caption{Same as Figure~\ref{apdx:sys:fig:sdepth} but for the mean
  airmass. 
  \label{apdx:sys:fig:airmass}}
\end{figure}

\begin{figure}
\begin{center}
\includegraphics[width=82mm]{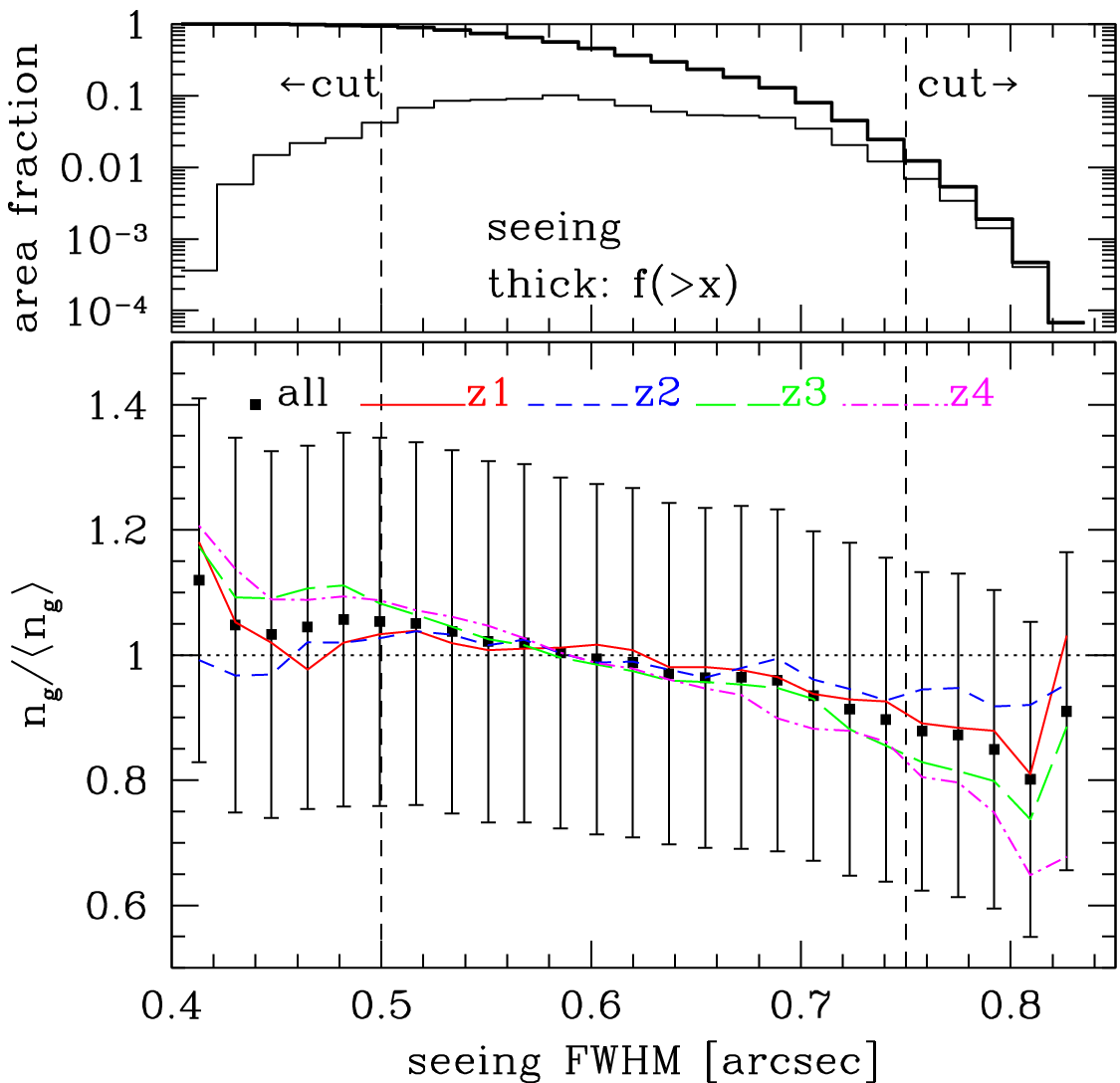}
\end{center}
\caption{Same as Figure~\ref{apdx:sys:fig:sdepth} but for the seeing
  (full width half maximum).  
  \label{apdx:sys:fig:seeing}}
\end{figure}

\begin{figure}
\begin{center}
\includegraphics[width=82mm]{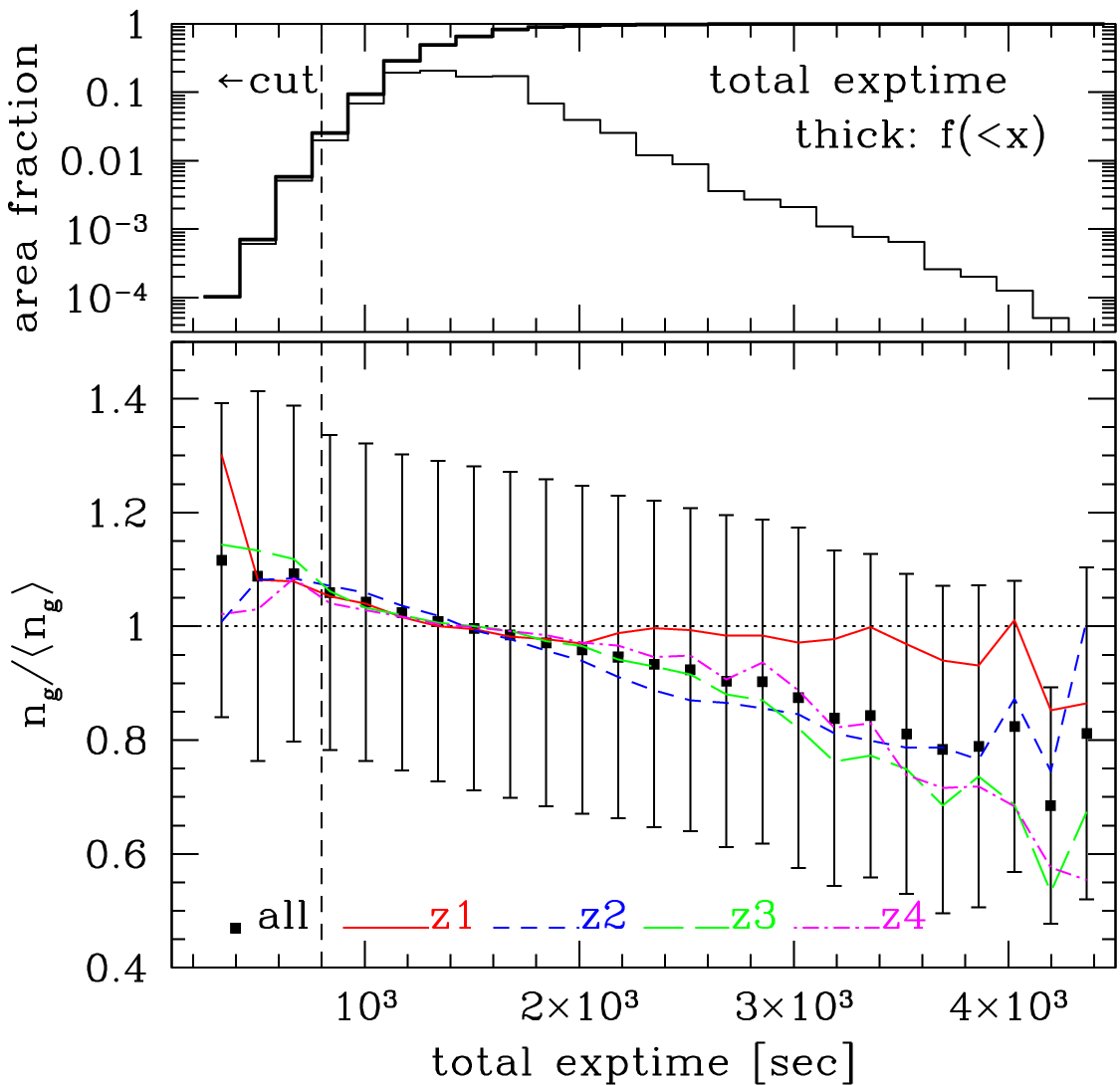}
\end{center}
\caption{Same as Figure~\ref{apdx:sys:fig:sdepth} but for the total
  exposure time.  
  \label{apdx:sys:fig:exptime}}
\end{figure}

\begin{figure}
\begin{center}
\includegraphics[width=82mm]{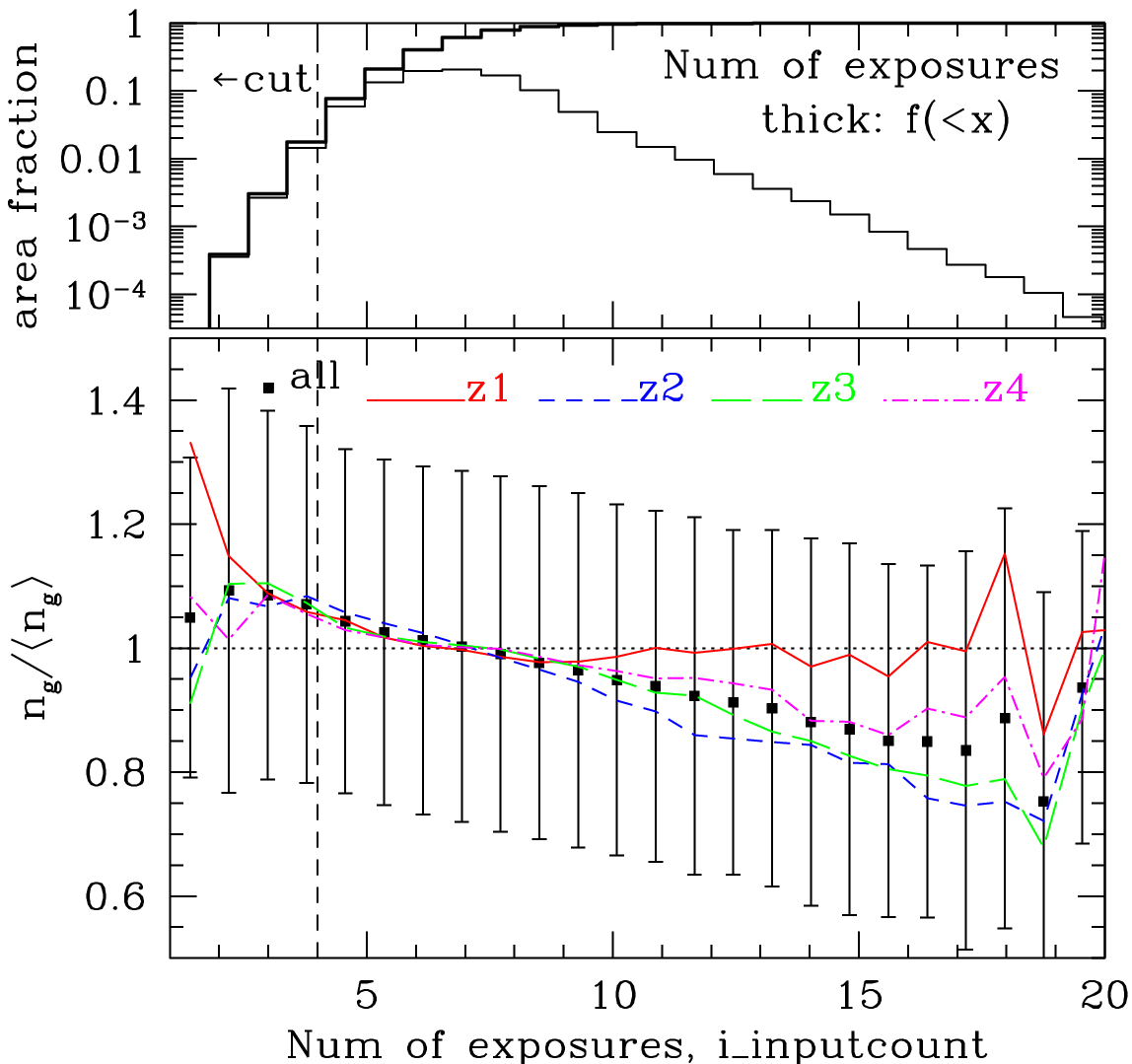}
\end{center}
\caption{Same as Figure~\ref{apdx:sys:fig:sdepth} but for the number of
  input frames, from the databese quantity {\tt i\_inputcount}.  
  \label{apdx:sys:fig:gicount}}
\end{figure}

\begin{figure}
\begin{center}
\includegraphics[width=82mm]{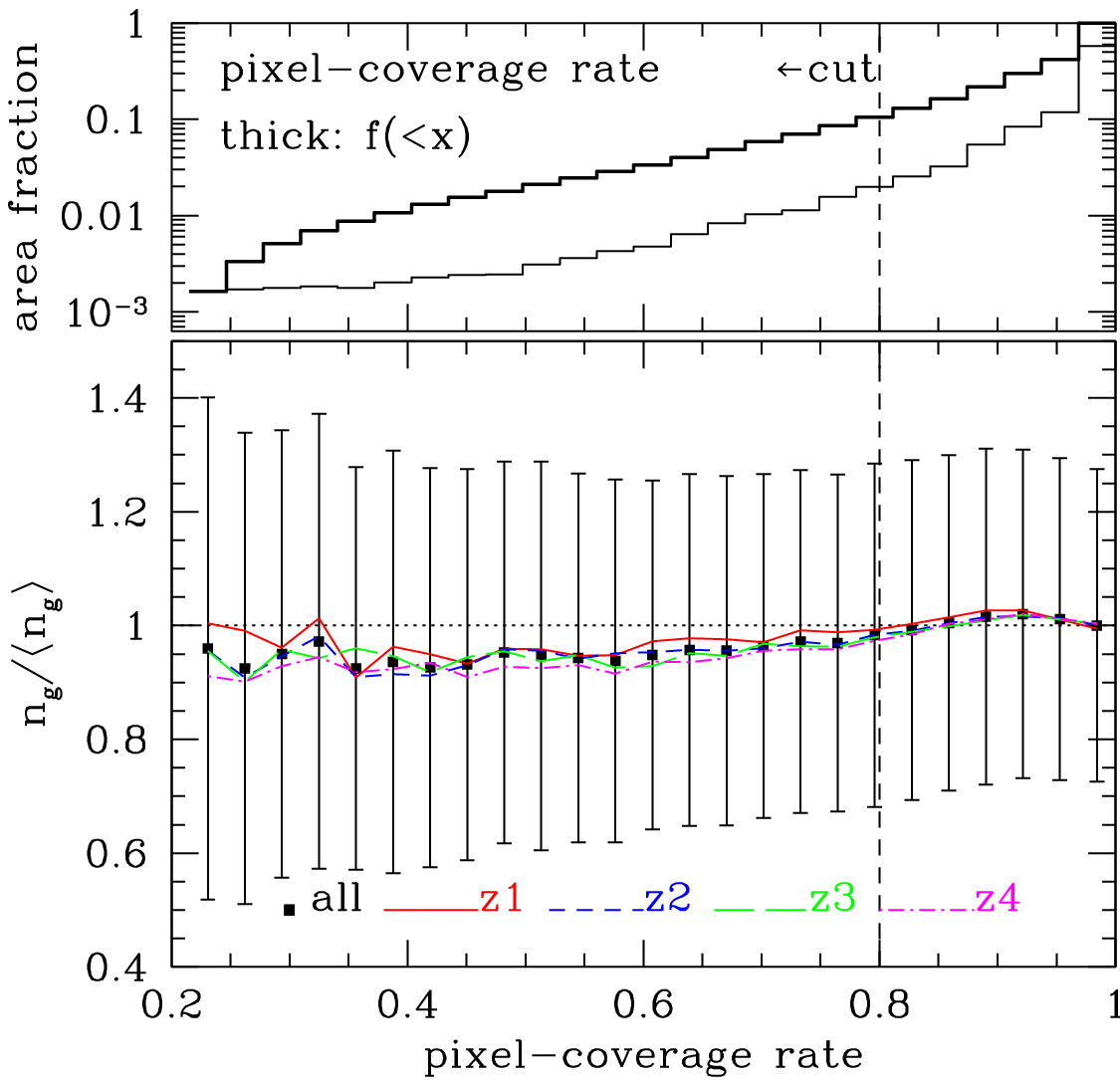}
\end{center}
\caption{Same as Figure~\ref{apdx:sys:fig:sdepth} but for the
  {\tt pixel-coverage rate}.  
  \label{apdx:sys:fig:pixfill}}
\end{figure}

Here, we describe details of area cuts of galaxy density maps to
mitigate effects of systematics on galaxy clustering measurements.
Our cuts are based on the inspection of the relation between a measured
quantity of systematics and the mean galaxy number density.
We compute the mean galaxy number density as a function of a measured
quantity of systematics.
The resulting plots are shown in
Figures~\ref{apdx:sys:fig:sdepth}-\ref{apdx:sys:fig:pixfill}.
We inspect the plots and determine threshold values for area cuts.
Our aim of this step is to trim the regions where systematics significantly
affect galaxy distributions.
Therefore, correlations between some measured quantities of systematics
and the mean galaxy density may remain after those cuts, which are
mitigated in the next step, the contamination deprojection described in
Section~\ref{sssec:mitigating}. 

In Figures~\ref{apdx:sys:fig:sdepth}-\ref{apdx:sys:fig:pixfill},
top panel shows the area fraction as a function of a measured quantity of
systematic, whereas, bottom panel shows the mean galaxy number density
as a function of a binned systematics quantity normalized by the mean
galaxy number density over all the bins.
Error bars show the root-mean-square of galaxy number 
densities over pixels within each bin.
To make those plots, the galaxy density maps and the systematic maps
with {\tt HEALPix} resolution parameter $N_{\rm side}=2048$ are used. 
The size of the error bars is predominantly determined by galaxy
clustering, thus representing the amplitude of galaxy clustering on
$\sim 2$ arcmin scale, and therefore exhibits little dependence on the
$x$-axis value.
The threshold values are shown by vertical dashed lines.
The threshold values and fractions of removed area are summarized in
Table~\ref{table:systematics}.
In total, 29.5 percent of the survey area is removed.

Note that excluding some galaxies from the original catalog may alter
the redshift distributions of galaxies in four tomographic samples.
We check this point by comparing the stacked probability distribution
functions of the redshift (estimated by {\tt dNNz} method) of galaxy
samples with and without area cuts applied.
We found that the PDFs of these two are almost identical,
and thus we conclude that the area cuts does not significantly alter
the redshift distributions of galaxy samples (see section
\ref{sssec:deltazparams} for further details).

Below, we discus three cases that deserve particular
emphasis:
\begin{itemize}
\item The star number density shown in Figure~\ref{apdx:sys:fig:nstar}.
  A positive correlation between the star number density and the galaxy
  number density is clearly seen. The contamination of stars in the
  galaxy sample due to mis-classification is a possible explanation
  for this correlation, at least in the regions of higher star number
  density. However, the cause of the decline observed at the lowest range
  ($\lesssim 2/$pixel) is unknown, it is possible that a secondly
  parameter, which reduce both galaxy and star density, may be involved.
  Since the correlation curve is very smooth, we do not set a threshold
  for this case.
\item The seeing condition shown in Figure~\ref{apdx:sys:fig:seeing},
  where a negative correlation is clearly observed.
  We set both the lower and upper threshold for this.
  The upper threshold is applied due to the poor observing
  conditions, whereas the lower threshold was set because the slope of the
  correlation fluctuated erratically.
\item The total exposure time and number of exposures shown in
  Figures~\ref{apdx:sys:fig:exptime} and \ref{apdx:sys:fig:gicount},
  respectively.
  Those two essentially have the same information.
  The negative correlation seen in those Figures is due to
  the following two reasons: One is the positive correlation between total
  exposure time and seeing, which is indeed seen in the HSC-Y3
  shape catalog.
  The other is the resolution factor cut imposed in making HSC-Y3 shape
  catalog \citep{Li_2022}, the resolution factor is ratio between the size of
  PSF and the size of galaxy, and is used to quantify the extent to
  which the galaxy is resolved compared to the PSF.
  Due to this cut, smaller galaxies compared with the
  PSF size were removed from the catalog.
  The combination of the above two factors leads to the negative
  correlation between the total exposure time and the mean galaxy number
  density.  
\end{itemize}

%
%

%
%

%
%

%
%

%
%

%
%

%
%

%
%

%
%

%
%

\end{appendix}

\clearpage
\bibliographystyle{apj}

\end{document}